\documentclass[prd,preprint,aps,showpacs,tightenlines,eqsecnum]{revtex4}

\usepackage{amsmath}
\usepackage{amssymb}

\newcommand{\beq}{\begin{equation}}
\newcommand{\eeq}{\end{equation}}
\newcommand{\beqa}{\begin{eqnarray}}
\newcommand{\eeqa}{\end{eqnarray}}

\begin{document}

\title{Physical equivalence between
 the covariant and physical graviton two-point functions in de~Sitter spacetime}
\author{Mir Faizal}
\email{faizal.mir@durham.ac.uk}
\affiliation{Centre for Particle Theory,
Department of Mathematical Sciences, Durham University, South Road,
Durham DH1 3LE, United Kingdom}
\author{Atsushi Higuchi}
\email{atsushi.higuchi@york.ac.uk}
\affiliation{Department of Mathematics, University of York,
Heslington, York YO10 5DD, United Kingdom}
\date{July 31, 2012}


\begin{abstract}
It is known that the covariant graviton two-point function in de~Sitter spacetime is infrared divergent for some choices of gauge parameters. On the other hand it is also known that there are no infrared divergences requiring an infrared cutoff
for the physical graviton two-point function for this spacetime in the transverse-traceless-synchronous gauge in the global coordinate system.
We show in this paper that the covariant graviton Wightman two-point function with two gauge parameters is equivalent to the physical one in the global coordinate system
in the sense that they produce the same two-point function of any local gauge-invariant tensor linear in the graviton field such as the linearized Weyl tensor.  This confirms the fact, pointed out decades ago, that the infrared divergences of the graviton two-point function in the covariant gauge for some choices of gauge parameters are gauge artifact in the sense that they do not contribute to the Wightman
two-point function of any local gauge-invariant tensor field in linearized theory.
\end{abstract}

\maketitle

\section{Introduction}

Infrared (IR) divergences of graviton two-point functions have been a matter of contention for over two decades.
There are two separate issues that are sometimes mistakenly thought to be related.  One issue is the IR divergences of
the physical graviton two-point function in the transverse-traceless-synchronous gauge in conformally-flat
coordinates~\cite{FordParker2,Allen86a,HiguchiCQG}. (See, e.g.\ Refs.~\cite{Higuchi:2011vw,Miao:2011ng}, for recent works on this issue.)
The other is the IR divergences of the covariant gauge for some choices of gauge parameters~\cite{AIT,AntoniadisMottola}.
(There is also the issue of large-distance growth of the two-point function, which will not be discussed in this paper.)
Since linearized gravity has gauge invariance, it is important to determine whether or not these IR divergences are gauge artifacts.  One of the reasons why the research community has not reached a consensus about this question seems to be that, when it is asserted that some IR divergences are a gauge artifact,
their precise definition is not made sufficiently clear.

The main purpose of this paper is to clarify in what sense the IR divergences of the graviton Wightman two-point function in the covariant gauge for some choices of gauge parameters are a gauge artifact.
(Below, by a two-point function we mean a Wightman two-point function unless otherwise stated.)  This is in fact an old result of
Allen~\cite{Allen86b}. We add to this result by showing that the covariant graviton two-point function with any choice of gauge parameters is physically equivalent to the physical one in the transverse-traceless-synchronous gauge in global coordinates~\cite{HiguchiWeeks}, which suffers no IR divergences.   This will also imply that the two-point function of any local gauge-invariant tensor field linear in the graviton field
evaluated in the covariant gauge is independent of gauge parameters as expected.

Miao, Tsamis and Woodard~\cite{MTW} find that the covariant two-point function corresponding to an \emph{IR-finite} choice of gauge parameters~\cite{Allen86b,HKcovariant}, the `strictly enforced' de~Donder gauge, is IR divergent in the Poincar\'e patch of de~Sitter spacetime, which is the spatially-flat expanding half of this spacetime.  We confirm, however, that IR divergences of the two-point function for a tachyonic scalar field, which is partly responsible for the breaking of de~Sitter invariance in Ref.~\cite{MTW}, are absent in global de~Sitter spacetime. We also find no IR divergences in the tensor sector of the
two-point function.  Thus, the covariant two-point function constructed using the mode-sum method agrees with the IR-finite two-point function in the Euclidean approach also in the de~Donder gauge.  (This gauge should probably be avoided in perturbation theory in any case because the corresponding two-point function behaves rather badly at large separation.)

We emphasize that this paper has nothing to say about interacting theory. In particular we do not couple the covariant graviton two-point function even to an external stress-energy tensor field.  Thus, in this paper the covariant two-point function is regarded as a graviton correlator and is shown to be equivalent to the physical one in Ref.~\cite{HiguchiWeeks} as such.  If the gravitons are coupled to an external stress-energy tensor, for example, there will be nonlocal interaction terms in the physical gauge of Ref.~\cite{HiguchiWeeks} similar to the Coulomb-interaction term in QED in the Coulomb gauge (see, e.g.\ Ref.~\cite{BjorkenDrell}), and an explicit demonstration of the equivalence between the physical and covariant gauges would be rather nontrivial.

In linearized gravity the two-point function of the graviton field $h_{ab}(x)$
has no physical meaning by itself because this theory has gauge invariance under the gauge transformation,
\beq
\delta h_{ab}(x) = \nabla_a\Lambda_b(x) + \nabla_b \Lambda_a(x),
\eeq
where $\Lambda_a(x)$ is any vector field.  Here,
the covariant derivative is the one compatible with the background de~Sitter metric, $g_{ab}(x)$.
One can find tensor fields at $x$ that are linear in $h_{ab}$ and are invariant under this gauge
transformation. An example of such a tensor field is
the linearized Weyl tensor
$W_{abcd}(x)  =  \widetilde{W}_{[ab][cd]}(x)$,
where
\beq
\widetilde{W}_{abcd}(x)  =  \nabla_{c} \nabla_{b}h_{ad}(x) + H^2g_{ad}(x)h_{cb}(x).
\eeq
Here, the constant $H$ is the Hubble constant of de~Sitter spacetime. (See Ref.~\cite{StewartWalker} for conditions for a local
tensor field to be gauge invariant.)  The two-point function of $W_{abcd}(x)$ evaluated in the covariant gauge
can be found in Ref.~\cite{Spyros} (with a corrigendum to be published).

Now, suppose that a graviton two-point function
$\Delta_{aba'b'}(x,x') = \langle 0|h_{ab}(x)h_{a'b'}(x')|0\rangle$ can be written as
\beq
\Delta_{aba'b'}(x,x') = \tilde{\Delta}_{aba'b'}(x,x')+
\nabla_{(a}Q_{b)a'b'}(x,x') + \nabla_{(a'}Q_{|ab|b')}(x,x'), \label{equivalent}
\eeq
for some $Q_{aa'b'}(x,x')$ and $Q_{aba'}(x,x')$. (In this paper we use the convention of Ref.~\cite{AllenJacobson} that primed indices are associated with point $x'$ and unprimed indices with point $x$.)  Then the two-point function of a local gauge-invariant tensor field
linear in $h_{ab}$ will be the same whether one uses $\Delta_{aba'b'}(x,x')$ or $\tilde{\Delta}_{aba'b'}(x,x')$ as the
graviton two-point function.  This motivates the following definition: we say that the two graviton two-point functions,
$\Delta_{aba'b'}(x,x')$ and $\tilde{\Delta}_{aba'b'}(x,x')$, are \emph{physically equivalent in linearized gravity} if
Eq.~(\ref{equivalent}) is satisfied for some $Q_{aa'b'}(x,x')$ and $Q_{aba'}(x,x')$, which are not required to be bounded.

A more precise formulation of the graviton two-point function would correspond to defining it in the smeared form as
\beq
D(f^{(1)},f^{(2)}) = \int d^4x\sqrt{-g(x)} \int d^4x'\sqrt{-g(x')} f^{(1)ab}(x)f^{(2)a'b'}(x')
\Delta_{aba'b'}(x,x'),  \label{characterize}
\eeq
where $f^{(1)ab}(x)$ and $f^{(2)a'b'}(x')$ are smooth, compactly-supported and divergence-free
symmetric tensor fields in de~Sitter spacetime.
Thus, the two-point function $D$ would be defined as a functional on the space of pairs of
smooth, compactly-supported and divergence-free symmetric tensor fields.
In such a definition, the functions
$\Delta_{aba'b'}(x,x')$ and $\tilde{\Delta}_{aba'b'}(x,x')$ satisfying Eq.~(\ref{equivalent}) can be regarded as two
representatives of the same two-point function $D$. [It can be shown that there are `sufficiently many' smooth,
compactly-supported and divergence-free symmetric tensor fields for characterizing the gauge-invariant content of the graviton
two-point function as in Eq.~(\ref{characterize}).]

Now suppose that a graviton two-point function $\Delta_{abab'}(x,x')$
has an IR cutoff $\epsilon$ and that it is divergent in the limit
$\epsilon\to 0$.  If it is physically equivalent in linearized gravity to $\Delta_{aba'b'}(x,x')$ that is not IR
divergent, then the two-point function of a local gauge-invariant tensor field will not depend on $\epsilon$, i.e.\ will
not be IR divergent.  What we show in this paper is that the covariant graviton two-point function
for any
choice of gauge parameters is physically equivalent in linearized gravity to the graviton two-point function in
the transverse-traceless-synchronous gauge in global coordinates,
which is IR finite~\cite{HiguchiWeeks}.  This will
imply that the IR divergences of the covariant two-point function for a certain gauge choice can be said to be
a gauge artifact
in linearized gravity in the sense that the divergences will not manifest themselves in the
two-point function of any local gauge-invariant tensor field linear in the graviton field,
confirming and clarifying
the claim in Ref.~\cite{Allen86b}.

The rest of the paper is organized as follows.  In Sec.~\ref{freefields} we summarize some properties of the solutions to the free field equations that we will need later
for scalar, vector and symmetric tensor fields in de~Sitter spacetime.  We leave the explicit expressions of these solutions to Appendix~\ref{explicit-solutions}.
In Sec.~\ref{physical} we review the physical two-point function
in the transverse-traceless-synchronous gauge in global coordinates.
In Sec.~\ref{solutions-covariant} we find all solutions to the field equation in the covariant gauge with two parameters.
In Sec.~\ref{innerproduct} we describe the quantization of
linearized gravity in the covariant gauge in de~Sitter spacetime. Then we construct the covariant
two-point function using the mode-sum method and show that it is equivalent to the physical two-point function
of Ref.~\cite{HiguchiWeeks}. In Sec.~\ref{summary} we summarize the results in
this paper.
We give explicit expressions for solutions to the free-field equations
in Appendix~\ref{explicit-solutions}. Appendix~\ref{ortho-appendix} contains a technical result used in Sec.~\ref{innerproduct}.
In Appendix~\ref{app:MTWrebuttal} the scalar two-point function, including the tachyonic case, is constructed by the mode-sum method in global de~Sitter spacetime.  We also show how this IR-finite two-point function can be recovered in Poincar\'e patch by subtracting the IR divergences.
In Appendix~\ref{consistency} we explicitly show that the covariant two-point function constructed in this paper is the same as that obtained in the Euclidean approach~\cite{HKcovariant} for spacelike-separated points.
We use the metric signature $-+++$ and let $\hbar=c= 1$  and take the metric of de~Sitter spacetime to be
\beq
ds^2 = - dt^2 + \cosh^2 t d\Omega^2, \label{dSmetric}
\eeq
where $d\Omega^2$ is the line element on the unit $3$-sphere ($S^3$), throughout this paper.
Thus, we choose units such that the Hubble constant is $1$.  A point $x$ in this spacetime has
coordinates $(t,\mathbf{x})$, where $\mathbf{x}$ is a point on $S^3$.

\section{Solutions to free-field equations} \label{freefields}

In this section we summarize some known properties of the solutions to the free-field equations for spin
$0$, $1$ and $2$ of arbitrary mass in de~Sitter spacetime following Ref.~\cite{HiguchiJMP}.  We present the
explicit solutions in Appendix~\ref{explicit-solutions}.
First we recall that the scalar, transverse vector and transverse-traceless tensor spherical harmonics on $S^3$, which we denote by $Y^{(0\ell\sigma)}$, $Y_i^{(1\ell\sigma)}$, and $Y_{ij}^{(2\ell\sigma)}$,
are orthonormal eigenfunctions of the Laplace-Beltrami operator $\tilde{\nabla}^2$ on $S^3$ satisfying
\beqa
- \tilde{\nabla}^2 Y^{(0\ell\sigma)} = \ell(\ell+2)Y^{(0\ell\sigma)},\,\,\ell=0,1,2,\ldots,\\
-\tilde{\nabla}^2 Y^{(1\ell\sigma)}_i = \left[\ell(\ell+2)-1\right]Y_i^{(1\ell\sigma)},\,\,\ell=1,2,3,\ldots,\\
-\tilde{\nabla}^2 Y^{(2\ell\sigma)}_{ij} = \left[\ell(\ell+2)-2\right]Y_{ij}^{(2\ell\sigma)},\,\,\ell=2,3,4,\ldots,
\eeqa
where $\sigma$ represents all labels other than $\ell$ (see, e.g.\ Refs.~\cite{ChodosMyers,RubinOrdonez}).

Let us start with the solutions to the scalar field equation,
\beq
\left( -\Box + \mu^2\right)\phi = 0. \label{scalareq}
\eeq
(The solutions we present here are valid for $\mu^2>0$ and for most negative values of $\mu^2$.)
We can choose the solutions to be proportional to $Y^{(0\ell\sigma)}$.
We denote the `positive-frequency' solutions that determine the Bunch-Davies (or Euclidean) vacuum~\cite{GibbonsHawking,BunchDavies} proportional to $Y^{(0\ell\sigma)}$
by $\phi^{(\mu^2;\ell\sigma)}(x)$. (We mean by `positive-frequency' solutions the coefficient functions of annihilation operators when the field is quantized.)  They are
\beq
\phi^{(\mu^2;\ell\sigma)}(x) \propto (\cosh t)^{-1}{\rm P}_{L_0+1}^{-(\ell+1)}
(i\sinh t)Y^{(0\ell\sigma)}(\mathbf{x}), \label{scalarmodes}
\eeq
with $L_0 = -\frac{3}{2}+\sqrt{\frac{9}{4}-\mu^2}$,
where ${\rm P}_{L_0+1}^{-(\ell+1)}(z)$ are the Legendre functions of the first kind given in terms of
Gauss's hypergeometric function as
\beq
{\rm P}_{L_0+1}^{-(\ell+1)}(z)=\frac{1}{(\ell+1)!}\left(\frac{1-z}{1+z}\right)^{(\ell+1)/2}
F\left(-L_0-1,L_0+2;\ell+2;\frac{1-z}{2}\right).
\eeq
These solutions and their complex conjugates, $\overline{\phi^{(\mu^2;\ell\sigma)}}$,
form a complete set of solutions to Eq.~(\ref{scalareq}).

We define the Klein-Gordon inner product for two solutions $\phi^{(1)}$ and $\phi^{(2)}$ to Eq.~(\ref{scalareq}) as follows:
\beq
\langle \phi^{(1)},\phi^{(2)}\rangle_{\rm KG} = i\int_{\Sigma}d\Sigma_a
\left[\overline{\phi^{(1)}}\nabla^a \phi^{(2)} - (\nabla^a\overline{\phi^{(1)}})\phi^{(2)}\right],
\label{scalarKG}
\eeq
where $d\Sigma_a = d\Sigma n_a$ with $n^a$ being
the future pointing unit normal vector to the Cauchy surface $\Sigma$.
We normalize the solutions $\phi^{(\mu^2;\ell\sigma)}$ by requiring
\beq
\langle \phi^{(\mu^2;\ell\sigma)},\phi^{(\mu^2;\ell'\sigma')}\rangle_{\rm KG}
= \delta^{\ell\ell'}\delta^{\sigma\sigma'}.
\eeq
The orthogonality follows from that of the spherical harmonics $Y^{(0\ell\sigma)}$ on $S^3$.  We also note that
$\phi^{(\mu^2;\ell\sigma)}$ are orthogonal to $\overline{\phi^{(\mu^2;\ell'\sigma')}}$ with respect to the
Klein-Gordon inner product.

We write the field equation for a transverse vector field $A_a$ satisfying $\nabla_a A^a = 0$ as
\beq
\left(-\Box +3+\mu^2\right)A_a = 0. \label{vectoreq}
\eeq
The gauge invariant equation, $\nabla^b(\nabla_aA_b - \nabla_bA_a) = 0$, is equivalent to
$(-\Box +3)A_a = 0$, i.e.\ the $\mu=0$ case of Eq.~(\ref{vectoreq}).  This can readily be seen by
recalling that $R_{abcd} = g_{ac}g_{bd} - g_{ad}g_{bc}$.  We will be particularly
interested in the case $\mu^2 = -6$, which is equivalent to $\nabla^b(\nabla_aA_b + \nabla_b A_a)=0$.

There are two classes of solutions to Eq.~(\ref{vectoreq}).  We introduce a label $m$ to distinguish between these classes.  The `positive-frequency' solutions will be denoted $A^{(\mu^2;m\ell\sigma)}_a$.
Those with $m=0$ have the time component given by
\beq
A_0^{(\mu^2;0\ell\sigma)} \propto (\cosh t)^{-2}{\rm P}_{L_1 + 1}^{-(\ell+1)}(i\sinh t) Y^{(0\ell\sigma)}(\mathbf{x}),
\,\,\ell\geq 1,
\eeq
where
\beq
L_1 = - \frac{3}{2} + \sqrt{\frac{1}{4} - \mu^2}.
\eeq
The space components, $A_i^{(\mu^2;0\ell\sigma)}$, are obtained by postulating
$A_i^{(\mu^2;0\ell\sigma)} = f_\ell(t) \tilde{\nabla}_i Y^{(0\ell\sigma)}$,
where $\tilde{\nabla}_i$ is the covariant derivative on $S^3$, and solving the equation
$\nabla^a A_a = 0$ for $f_\ell(t)$. (This equation cannot be solved for $\ell=0$.  Hence there are no solutions with $\ell=0$.)  The solutions with $m=1$ have $A_0^{(\mu^2;1\ell\sigma)}=0$ and
\beq
A_i^{(\mu^2;1\ell\sigma)} \propto {\rm P}_{L_1+1}^{-(\ell+1)}(i\sinh t) Y_i^{(\ell\sigma)}(\mathbf{x}),\,\,\,\ell\geq 1.
\eeq.

We define the Klein-Gordon inner product for two transverse solutions $A^{(1)}_a$ and $A^{(2)}_a$ to Eq.~(\ref{vectoreq}) as
\beq
\langle A^{(1)}, A^{(2)}\rangle_{\rm KG} = i\int_{\Sigma}d\Sigma_a
\left[ \overline{A^{(1)b}}\nabla^a A^{(2)}_b - (\nabla^a \overline{A^{(1)b}})A_b^{(2)}\right].
\eeq
Any two solutions with different sets of quantum numbers $m$, $\ell$ and $\sigma$ are orthogonal to each other with respect to this inner product. For $-6< \mu^2 < 0$ the `positive-frequency'
solutions with $m=0$ have negative norm with respect to this inner product whereas the $m=1$ solutions have
positive norm.  We normalize these solutions for $\ell\geq 2$ by requiring
\beq
\langle A^{(\mu^2;m\ell\sigma)},A^{(\mu^2;m'\ell'\sigma')}\rangle_{\rm KG}
= (-1)^{m+1}\delta^{mm'}\delta^{\ell\ell'}\delta^{\sigma\sigma'}.
\eeq
The solutions $A_a^{(\mu^2;m,\ell=1,\sigma)}$ become Killing vectors in the limit $\mu^2 \to -6$.  This implies that the Klein-Gordon inner product vanishes for these solutions because if $\xi^{(1)a}$ and $\xi^{(2)a}$ are Killing vectors, then
\beqa
\int_\Sigma d\Sigma_a (\xi^{(1)b}\nabla^a \xi^{(2)}_b - \xi^{(2)b}\nabla^a \xi^{(1)}_b)
& = & \int_\Sigma d\Sigma_a \nabla_b(\xi^{(1)a}\xi^{(2)b} - \xi^{(1)b}\xi^{(2)a})\nonumber \\
& = & 0
\eeqa
by the generalized Stokes theorem, which states that for any anti-symmetric tensor $F^{ab}$
\beq
\int_\Sigma d\Sigma_a \nabla_b F^{ab} = 0.
\eeq
For this reason we normalize the $\ell=1$ solutions as
\beq
\langle A^{(\mu^2;m,\ell=1,\sigma)},A^{(\mu^2;m',\ell=1,\sigma')}\rangle_{\rm KG}
=  (-1)^{m+1}(\mu^2+6)\delta^{mm'}\delta^{\sigma\sigma'}. \label{A1L1}
\eeq

We write the field equation for a transverse-traceless tensor field $H_{ab}$ satisfying $\nabla^bH_{ab} = 0$ and ${H^a}_a =0$ as
\beq
\left(-\Box +2+M^2\right) H_{ab} = 0.  \label{tensoreq}
\eeq
The $M=0$ case corresponds to linearized gravity.
There are three classes of solutions distinguished by the label $m =0,1,2$.  We write the `positive-frequency' solutions as $H^{(M^2;m\ell\sigma)}_{ab}$.  Those with $m=0$ have
\beq
H_{00}^{(M^2;0\ell\sigma)}\propto (\cosh t)^{-3}{\rm P}_{L_2+1}^{-(\ell+1)}(i\sinh t)Y^{(0\ell\sigma)}(\mathbf{x}),
\,\,\,\ell\geq 2, \label{0comp}
\eeq
where
\beq
L_2  =  -\frac{3}{2} + \sqrt{\frac{9}{4} - M^2}.
\eeq
The other components are obtained by postulating that
$H_{0i} = f_\ell(t)\tilde{\nabla}_i Y^{(0\ell\sigma)}$,
$H_{ij} = g^{(1)}_\ell(t)\tilde{\nabla}_i\tilde{\nabla}_j Y^{(0\ell\sigma)}
+ g^{(2)}_\ell(t)\tilde{\eta}_{ij}Y^{(0\ell\sigma)}$,
where $\tilde{\eta}_{ij}$ is the metric on $S^3$, and solving the equations
$\nabla^b H_{ab}=$ and ${H^a}_a = 0$
for the functions $f_\ell(t)$, $g^{(1)}_\ell(t)$ and $g^{(2)}_\ell(t)$.
This is not possible if $\ell=0$ or $1$ in Eq.~(\ref{0comp}).
The `positive-frequency' solutions with $m=1$ have $H_{00}^{(M^2;1\ell\sigma)} = 0$ and
\beq
H_{0i}^{(M^2;1\ell\sigma)} \propto
(\cosh t)^{-1}{\rm P}_{L_2+1}^{-(\ell+1)}(i\sinh t)Y^{(1\ell\sigma)}_i(\mathbf{x}),\,\,\,\ell\geq 2.
\label{0icomp}
\eeq
Then we postulate that $H_{ij} = f_\ell(t)\tilde{\nabla}_{(i}Y_{j)}^{(1\ell\sigma)}$ and solve
$\nabla^a H_{ab} = 0$ for $f_\ell(t)$. This is not possible if $\ell=1$ in
Eq.~(\ref{0icomp}).  Finally, the `positive-frequency' solutions with
$m=2$ have $H_{00}^{(M^2;2\ell\sigma)}= H_{0i}^{(M^2;2\ell\sigma)} = 0$ and
\beq
H_{ij}^{(M^2;2\ell\sigma)} \propto \cosh t\,{\rm P}_{L_2+1}^{-(\ell+1)}(i\sinh t)Y^{(2\ell\sigma)}_{ij}(\mathbf{x}),
\,\,\,\ell\geq 2.
\eeq

We define the Klein-Gordon inner product for two transverse-traceless solutions $H_{ab}^{(1)}$ and $H_{ab}^{(2)}$ to Eq.~(\ref{tensoreq}) as
\beq
\langle H^{(1)}, H^{(2)}\rangle_{\rm KG} = i\int_{\Sigma}d\Sigma_a
\left[ \overline{H^{(1)bc}}\nabla^a H^{(2)}_{bc} - (\nabla^a \overline{H^{(1)bc}})H_{bc}^{(2)}\right].
\label{KGH}
\eeq
We can normalize the $m=2$ solutions as
\beq
\langle H^{(M^2;2\ell\sigma)},H^{(M^2;2\ell'\sigma')}\rangle_{\rm KG} =  2\delta^{\ell\ell'}\delta^{\sigma\sigma'}.\label{H1}
\eeq
The factor of $2$ here is for later convenience.
For $M=0$, i.e.\ for linearized gravity, Eq.~(\ref{tensoreq}) is satisfied by
$H_{ab} = \nabla_{(a} A_{b)}^{(-6;m\ell\sigma)}$.
Indeed one finds, using the associated Legendre equation (\ref{associated}) and the lowering and raising
differential operators, Eqs.~(\ref{lowering}) and (\ref{raising}),
\beq
H_{ab}^{(0;m\ell\sigma)} = \nabla_a A_b^{(-6;m\ell\sigma)} + \nabla_b A_a^{(-6;m\ell\sigma)},\,\,\,m=0,1,\,\,
\ell\geq 2, \label{non-indep1}
\eeq
after choosing a phase factor for $H_{ab}^{(0;m\ell\sigma)}$ appropriately~\cite{HiguchiSO4}.
Now, if $H^{(1)}_{ab}$ is any solution to Eq.~(\ref{tensoreq}) with $M^2=0$ and if
$H^{(2)}_{ab} = \nabla_{(a}A^{(2)}_{b)}$ with $(\Box + 3)A^{(2)}_a = 0$ so that $H^{(2)}_{ab}$ is a solution
to Eq.~(\ref{tensoreq}), then we find
\beqa
\langle H^{(1)},H^{(2)}\rangle_{\rm KG} & = &
i\int_{\Sigma}d\Sigma_a
\nabla_b\left[\overline{H^{(1)bc}}\nabla^aA^{(2)}_c - \overline{H^{(1)ac}}\nabla^bA_c^{(2)}
+ (\nabla^b\overline{H^{(1)ac}}-\nabla^a\overline{H^{(1)bc}})A_c^{(2)}\right]\nonumber \\
& = & 0
\eeqa
by the generalized Stokes theorem.  Thus,
\beq
\langle H^{(0;m\ell\sigma)},H^{(0;m'\ell'\sigma')}\rangle_{\rm KG} = 0,\,\,\ m,m'=0,1.
\eeq
It is also known that the solutions
$H^{(M^2;0\ell\sigma)}_{ab}$ have negative norm if $0 < M^2 <2$~\cite{Higuchiunp,Higuchi1987} whereas the solutions $H^{(M^2;1\ell\sigma)}$ have positive norm if $M^2>0$.  For these reasons, and since we will be interested in the $M\to 0$ limit,
we normalize the solutions with $m=0,1$ as
\beq
\langle H^{(M^2;m\ell\sigma)},H^{(M^2;m'\ell'\sigma')}\rangle_{\rm KG}  =  (-1)^{m+1}2M^2 \delta^{mm'}\delta^{\ell\ell'}\delta^{\sigma\sigma'}.\label{H2}
\eeq

\section{Physical graviton two-point function} \label{physical}

The Lagrangian for free gravitons in de~Sitter spacetime can be written as
\beqa
{\cal L}_{\rm inv} & = & \sqrt{-g}\left[ \frac{1}{2}\nabla_a h^{ac}\nabla^b h_{bc}
- \frac{1}{4}\nabla_a h_{bc}\nabla^a h^{bc} + \frac{1}{4}
\left(\nabla^a h - 2\nabla^b{h^a}_b\right)\nabla_a h\right. \nonumber \\
&& \,\,\,\,\,\, \left.
- \frac{1}{2}\left(h_{ab}h^{ab} + \frac{1}{2}h^2\right)\right] \label{Linvariant}
\eeqa
with $h \equiv {h^a}_a$.  The corresponding field equation is
\beqa
L_{ab}^{({\rm inv})cd}h_{cd} & \equiv & \frac{1}{2}\left[ -\Box h_{ab} + \nabla_a\nabla_c {h^c}_b
+ \nabla_b \nabla_c{h^c}_a - \nabla_a\nabla_b h\right.  \nonumber \\
&&\left. + g_{ab}\Box h - g_{ab}
\nabla_c\nabla_d h^{cd}\right] +  h_{ab} + \frac{1}{2} g_{ab}h = 0.  \label{ffeq}
\eeqa
It is well known that the gauge degrees of freedom can be used to impose the conditions
$\nabla^b h_{ab} = 0$ and $h=0$ (see, e.g.\ Ref.~\cite{HiguchiSO4}).
Then Eq.~(\ref{ffeq}) becomes
\beq
(\Box - 2)h_{ab} = 0.  \label{simpleeq}
\eeq
This equation is Eq.~(\ref{tensoreq}) with $M=0$.  Thus, its solutions are given by Eqs.~(\ref{firsteq})-(\ref{lasteq}) with $M=0$.

We have seen that the solutions $H_{ab}^{(0;m\ell\sigma)}$, $m=0,1$, are gauge solutions [see Eq.~(\ref{non-indep1})].
Hence only the solutions
$H_{ab}^{(0;2\ell\sigma)}$ represent physical excitations.  Retaining only these solutions corresponds to the synchronous
transverse-traceless gauge, $h_{0a}=0$, $\tilde{\nabla}_j{h^j}_i = 0$ and ${h^i}_i = 0$.
Quantization of the field $h_{ab}$ in this gauge, which we sometimes call the physical gauge,
proceeds as follows.  First we find the
canonical conjugate momentum current $p^{abc}$ as
\beqa
p^{abc} & = & \frac{1}{\sqrt{-g}}\frac{\partial\mathcal{L}}{\partial \nabla_a h_{bc}}\nonumber \\
& = & - \frac{1}{2}\nabla^a h^{bc},  \label{pdef}
\eeqa
where we have used the conditions $\nabla^b h_{ab}=0$ and ${h^a}_a=0$.
We note that the field equation (\ref{ffeq}) can be given as
\beq
\nabla_a p^{abc} - \frac{1}{\sqrt{-g}}\frac{\partial\mathcal{L}_{\rm inv}}{\partial h_{bc}} = 0.
\label{pfieldeq}
\eeq
We define the symplectic product between two solutions $h^{(1)}_{ab}$ and $h^{(2)}_{ab}$
to this equation as follows:
\beq
(h^{(1)},h^{(2)})_{\rm symp} = -i \int_\Sigma d\Sigma_a
(\overline{h^{(1)}_{bc}}p^{(2)abc} - \overline{p^{(1)abc}}h^{(2)}_{bc}), \label{symp1}
\eeq
where $p^{(1)abc}$ is obtained by substituting $h_{ab}=h_{ab}^{(1)}$ into Eq.~(\ref{pdef})
and similarly for $p^{(2)abc}$.
This product is independent of the Cauchy surface $\Sigma$ thanks to Eqs.~(\ref{pdef}) and
(\ref{pfieldeq}).  Then we find
\beqa
(H^{(0;2\ell\sigma)},H^{(0;2\ell'\sigma')})_{\rm symp} & = & \frac{1}{2}\langle H^{(0;2\ell\sigma)},H^{(0;2\ell'\sigma')}\rangle_{KG} \nonumber \\
&  = & \delta^{\ell\ell'}\delta^{\sigma\sigma'}.
\eeqa

We expand the quantum field $h_{ab}$
in the synchronous transverse-traceless gauge as
\beq
h_{ab}(x) = \sum_{\ell=2}^\infty \sum_{\sigma}\left[ a_{\ell\sigma}H_{ab}^{(0;2\ell\sigma)}(x) + a_{\ell\sigma}^\dagger \overline{H_{ab}^{(0;2\ell\sigma)}(x)}\right].
\eeq
We then impose the commutation relations $[a_{\ell\sigma},a_{\ell'\sigma'}^\dagger] = \delta_{\ell\ell'}\delta_{\sigma\sigma'}$ with all other commutators vanishing.  We define the Bunch-Davies
vacuum state $|0\rangle$ by requiring that $a_{\ell\sigma}|0\rangle=0$ for all $\ell$ and $\sigma$.  Then the
Wightman two-point function is readily found as
\beqa
\Delta_{aba'b'}^{({\rm phys})}(x,x') & = &
\langle 0|h_{ab}(x)h_{a'b'}(x')|0\rangle \nonumber \\
& = & \sum_{\ell=2}^\infty\sum_{\sigma} H_{ab}^{(0;2\ell\sigma)}(x)\overline{H_{a'b'}^{(0;2\ell\sigma)}(x')}.
\eeqa
This two-point function vanishes if any of the indices is `$0$'.  The space components can be found,
using Eq.~(\ref{physicalH}) with $M=0$, as
\beqa
\Delta_{iji'j'}^{({\rm phys})}(x,x') & = & \sum_{\ell=2}^\infty
(\ell-1)!(\ell+2)!\cosh t \cosh t'\,{\rm P}_1^{-(\ell+1)}(i\sinh t)
\overline{{\rm P}_1^{-(\ell+1)}(i\sinh t')}\nonumber \\
&& \times \sum_{\sigma} Y_{ij}^{(2\ell\sigma)}(\mathbf{x})\overline{Y_{i'j'}^{(2\ell\sigma)}}(\mathbf{x}'),
\label{phystwo}
\eeqa
where $x=(t,\mathbf{x})$ and $x'=(t',\mathbf{x}')$.
This series can be summed in a closed form. The result of this summation can be found
in Ref.~\cite{HiguchiWeeks}, in which it was shown that there are no IR divergences for this two-point function in the sense that it is well defined without any infrared cutoff.

\section{Solutions to the field equation in the covariant gauge} \label{solutions-covariant}

If we add a covariant gauge-fixing term in the Lagrangian, there will be solutions to the graviton field equations in addition to those given by Eqs.~(\ref{firsteq})-(\ref{lasteq}) with $M =0$.
In this section we describe all solutions including
these additional solutions to the graviton field equation in the covariant gauge.  These solutions will be used in the next section to find the two-point function.

The Lagrangian in the covariant gauge is
\beq
{\cal L} = {\cal L}_{\rm inv} + {\cal L}_{\rm gf},
\eeq
where ${\cal L}_{\rm inv}$ is given by Eq.~(\ref{Linvariant}) and where
\beq
{\cal L}_{\rm gf}  = - \frac{\sqrt{-g}}{2\alpha}
\left( \nabla_a h^{ab} - \frac{1+\beta}{\beta}\nabla^b h\right)
\left(\nabla^c h_{cb} - \frac{1+\beta}{\beta}\nabla_b h\right).
\eeq
We require that $\alpha>0$ for now.  We also assume $\beta > 0$, but most of our results
will be valid also for most negative values of $\beta$.
The Euler-Lagrange field equation derived from ${\cal L}_{\rm inv}+{\cal L}_{\rm gf}$ is
\beq
L_{ab}^{({\rm inv})cd}h_{cd} + L_{ab}^{({\rm gf})cd}h_{cd}=0, \label{E-L}
\eeq
where
$L_{ab}^{({\rm inv})cd}$ is defined by Eq.~(\ref{ffeq}) and where
\beqa
L_{ab}^{({\rm gf})cd}h_{cd} & = &
- \frac{1}{2\alpha}\left[\nabla_a\left(\nabla_c{h^c}_b - \frac{1+\beta}{\beta}\nabla_b h\right)
+ \nabla_b\left(\nabla_c{h^c}_a - \frac{1+\beta}{\beta}\nabla_a h\right)\right]\nonumber \\
&& + \frac{1+\beta}{\alpha\beta}g_{ab}\nabla_d\left(\nabla_c h^{cd} - \frac{1+\beta}{\beta}\nabla^d h\right).
\eeqa

Let us first find the solutions of the form
\beq
h^{(S)}_{ab} = \nabla_a\nabla_b B + g_{ab}\Psi.  \label{hs}
\eeq
By substituting this expression into Eq.~(\ref{E-L}) we find
\beq
\nabla_a\nabla_b X + g_{ab} Y = 0, \label{XYeq}
\eeq
where
\beqa
X & = & \frac{1}{\alpha\beta}\left[ (\Box -3\beta)B + (4 -\alpha\beta + 3\beta)\Psi\right],\\
Y & = & -\frac{1+\beta}{\alpha\beta^2}\Box(\Box - 3\beta)B + \left[1-\frac{4(1+\beta)^2}{\alpha\beta^2} + \frac{1+\beta}{\alpha\beta}\right]
\Box \Psi + 3\Psi.
\eeqa
This calculation is simplified by noting that $\nabla_a\nabla_b B$ does not contribute to
${L_{ab}}^{({\rm inv})cd}h_{cd}^{(S)}$ due to gauge invariance.
Eq.~(\ref{XYeq}) is obviously satisfied if $X=Y=0$.  These equations can be simplified by
solving the equation $X=0$ for $(\Box -3\beta)B$ and substituting the result into the equation $Y=0$.
Thus, the equations $X=Y=0$ can readily be shown to be equivalent to
\beqa
(\Box - 3\beta)B + \left[4-(\alpha - 3)\beta\right]\Psi & = & 0,\label{equation1} \\
(\Box -3\beta)\Psi & = & 0. \label{equation2}
\eeqa
The following solutions and their complex conjugates form a complete set of solutions to Eqs.~(\ref{equation1}) and
(\ref{equation2}):
\beqa
B^{(S1;\ell\sigma)} & = & \phi^{(3\beta;\ell\sigma)},\label{scalar1} \\
\Psi^{(S1;\ell\sigma)} & = & 0
\eeqa
and
\beqa
B^{(S2;\ell\sigma)} & = & -\left[4-(\alpha-3)\beta\right]\left.\frac{\partial\ }{\partial \mu^2}
\phi^{(\mu^2;\ell\sigma)}\right|_{\mu^2=3\beta},\\
\Psi^{(S2;\ell\sigma)} & = & \phi^{(3\beta;\ell\sigma)},\label{scalar2}
\eeqa
where $\partial/\partial\mu^2$ denotes the first derivative with respect to $\mu^2$ (rather than the second derivative with respect to $\mu$).
Eq.~(\ref{equation1}) can be verified for the solutions
$(B^{(S2;\ell\sigma)},\Psi^{(S2;\ell\sigma)})$ by noting
\beq
\left. \frac{\partial\ }{\partial\mu^2}
\left( -\Box +\mu^2\right)\phi^{(\mu^2;\ell\sigma)}\right|_{\mu^2=3\beta}=0.
\eeq
Note that the mass of these modes are $\beta$-dependent~\cite{Allen86b,HKcovariant}.  In particular,
if $\beta< 0$, then they are tachyonic because their mass squared is $\mu^2 = 3\beta < 0$.
Unfortunately, the familiar de~Donder gauge condition, $\nabla^b h_{ab} - \frac{1}{2}\nabla_a h = 0$,
corresponds to
$\mu^2 = 3\beta = -6$ (and $\alpha\to 0$).  Thus, these modes are tachyonic for the de~Donder
gauge~\cite{MTW,AllenTuryn}.  The gauge chosen by Antoniadis and Mottola~\cite{AntoniadisMottola},
$\nabla^b h_{ab} - \frac{1}{4}\nabla_a h = 0$, corresponds to $3\beta = -4$.  This choice has an additional
problem: the scalar field theory suffers IR divergences if $\mu^2 = -k(k+3)$ for $k=0,1,2,\ldots$~\cite{Allen86b}. [This fact can readily be seen from Eq.~(\ref{Euclscalar}).]
This is the cause of the IR divergences in the Antoniadis-Mottola gauge.
These problems can easily be avoided by requiring $\beta > 0$.

Although the de~Donder gauge ($3\beta=-6$) does not lead to IR divergences in the sense that the two-point function is finite without an IR cutoff, there are IR divergences in its expansion in terms
of momentum eigenfunctions in spatially-flat coordinate system~\cite{MTW}.  These divergences are due to the growth of the two-point function for large separation, which renders the momentum expansion ill-defined.  However, it is possible to regularize the IR-divergences in such a way that one recovers the finite two-point function when the regulator is removed as we show in Appendix~\ref{app:MTWrebuttal}.

Let us write the solutions to Eq.~(\ref{E-L}) corresponding to Eqs.~(\ref{scalar1})-(\ref{scalar2}) as
\beqa
S_{ab}^{(1;\ell\sigma)} & = & \nabla_a\nabla_b B^{(S1;\ell\sigma)}, \label{scalarsol1}\\
S_{ab}^{(2;\ell\sigma)} & = & \nabla_a\nabla_b B^{(S2;\ell\sigma)} + g_{ab}\Psi^{(S2;\ell\sigma)}.
\label{scalarsol2}
\eeqa
We show next that any solution $h_{ab}$ to Eqs.~(\ref{E-L}) can be decomposed as
$h_{ab} = h_{ab}^{(T)} + h^{(S)}_{ab}$, where $h^{(S)}_{ab}$ is a linear combination of the
solutions $S^{(A;\ell\sigma)}_{ab}$, $A=1,2$, and their complex conjugates and where
$\nabla^a\nabla^b h_{ab}^{(T)} = 0$ and $g^{ab}h^{(T)}_{ab}=0$.  For this purpose it is sufficient to show that
for any given solution $h_{ab}$ to Eq.~(\ref{E-L}) one can construct scalar fields
 $B$ and $\Psi$
satisfying Eqs.~(\ref{equation1}) and (\ref{equation2})
such that the field $h_{ab}^{(S)}= \nabla_a\nabla_b B + g_{ab}\Psi$ satisfies
$\nabla^ah_{ab} = \nabla^a h^{(S)}_{ab}$ and $g^{ab}h_{ab} = g^{ab}h^{(S)}_{ab}$.
To do so, for any solution $h_{ab}$ to Eq.~(\ref{E-L}) we define
\beq
\Phi(h)  = -\frac{1}{\alpha}\left(\nabla_a \nabla_b h^{ab} - \frac{1+\beta}{\beta}\Box h\right).
\eeq
Then by taking the divergence of Eq.~(\ref{E-L}) twice we find
\beq
\left(\Box - 3\beta\right)\Phi(h) =  0.  \label{EL1}
\eeq
This calculation is made easier by noting that the tensor field ${L_{ab}}^{({\rm inv}) cd}h_{cd}$ is
divergence-free (the background Bianchi identity).
Next, by taking the trace of Eq.~(\ref{E-L}) and using Eq.~(\ref{EL1}) we find
\beq
\left(\Box - 3\beta \right)h + [4-(\alpha-3)\beta]\Phi(h)=0. \label{EL2}
\eeq
Now, we define
\beq
h^{(S)}_{ab} = \nabla_a\nabla_bB(h) + g_{ab}\Psi(h),
\eeq
where
\beqa
B(h) & = & \frac{1}{3\beta}\left( h - \frac{\alpha-3}{3}\Phi(h)\right),\\
\Psi(h) & = & \frac{1}{3\beta}\Phi(h).
\eeqa
Then, one can readily see that Eqs.~(\ref{EL1}) and (\ref{EL2}) imply Eqs.~(\ref{equation1}) and (\ref{equation2}).  Thus, $h_{ab}^{(S)}$ is
a solution to Eq.~(\ref{E-L}).  Moreover, we find
\beqa
g^{ab}h^{(S)}_{ab} & = & \Box B(h) + 4\Psi(h) \nonumber \\
& = & h
\eeqa
and
\beq
\nabla^a\nabla^bh^{(S)}_{ab} - \frac{1+\beta}{\beta} g^{ab}\Box h^{(S)}_{ab}
 =  \nabla_a \nabla_b h^{ab} - \frac{1+\beta}{\beta}\Box h,
\eeq
and, hence,
\beq
\nabla^a\nabla^b h_{ab}^{(S)} = \nabla^a\nabla^b h_{ab}.
\eeq
Thus, any $h_{ab}$ satisfying Eq.~(\ref{E-L}) can be written
as $h_{ab} = h_{ab}^{(T)} + h_{ab}^{(S)}$, where $\nabla^a\nabla^b h^{(T)}_{ab} = 0$ and $g^{ab}h_{ab}^{(T)} = 0$.

Our next task is to construct all solutions to Eq.~(\ref{E-L}) satisfying $\nabla^a\nabla^b h_{ab}^{(T)}=0$ and $g^{ab}h_{ab}^{(T)} = 0$.  Eq.~(\ref{E-L}) becomes
\beqa
{L^{(T)}_{ab}}^{cd}h^{(T)}_{cd} & \equiv & -\frac{1}{2}\Box h_{ab}^{(T)} +\frac{1}{2}
\left( 1-\frac{1}{\alpha}\right)\left(\nabla_a\nabla_c {h^{(T)c}}_b
+ \nabla_b \nabla_c{h^{(T)c}}_a\right)+ h^{(T)}_{ab}\nonumber \\
& = & 0.  \label{massless}
\eeqa
We show that a complete set of solutions $h_{ab}^{(T)}$ is given by
\beqa
E_{ab}^{(1;m\ell\sigma)} & \equiv & H_{ab}^{(0;m\ell\sigma)},\,\,\,m=0,1,2,\,\,\,\ell\geq 2,
\label{tensor1}\\
E_{ab}^{(2;m\ell\sigma)} & \equiv & \lim_{M\to 0}
\frac{1}{M^2}\left( \nabla_a A_b^{(\mu^2; m\ell\sigma)} + \nabla_b A_a^{(\mu^2;m\ell\sigma)}
- H_{ab}^{(M^2;m\ell\sigma)}\right) \nonumber \\
& = & 2\alpha \left.\frac{\partial\ }{\partial \mu^2}
\nabla_{(a} A_{b)}^{(\mu^2;m\ell\sigma)}\right|_{\mu^2=-6} - \left.
\frac{\partial\ }{\partial M^2} H_{ab}^{(M^2;m\ell\sigma)}\right|_{M^2=0},\,\,\,m=0,1,\,\,\,\ell\geq 1,\nonumber \\
\label{tensor2}
\eeqa
and their complex conjugates,
with the identification
\beq
\mu^2 = \alpha M^2 - 6. \label{muM}
\eeq
We have defined $H_{ab}^{(M^2;m,\ell=1,\sigma)} = 0$ in the second equation.
The second equality in Eq.~(\ref{tensor2}) follows from Eq.~(\ref{non-indep1}), which is valid also
for $\ell=1$.

One can readily see that $h^{(T)}_{ab} = E^{(1;m\ell\sigma)}_{ab}$ and their complex conjugates give a complete set of solutions to Eq.~(\ref{massless}) under a stronger condition $\nabla^b h_{ab}^{(T)} = 0$.
The tensor fields $E^{(2;m\ell\sigma)}_{ab}$ and their complex conjugates are also solutions (under the
original condition $\nabla^a\nabla^bh_{ab}^{(T)}=0$) because
both $h_{ab}^{(M^2)} =
\nabla_{(a}A_{b)}^{(\mu^2;m\ell\sigma)}$ and $H_{ab}^{(M^2;m\ell\sigma)}$ are solutions to the
massive equation~\cite{FierzPauli}
\beq
L_{ab}^{(T)cd}h^{(M^2)}_{cd} + \frac{1}{2}M^2 (h^{(M^2)}_{ab}-g_{ab}h_c^{(M^2)c}) = 0. \label{massiveeq1}
\eeq
Then what is left to do is show that for any solution $h_{ab}^{(T)}$ of Eq.~(\ref{massless}) we can find
a linear combination $h_{ab}^{(W)}$ of $E^{(2;m\ell\sigma)}_{ab}$ and their complex conjugates such that
$C_a(h^{(T)}) \equiv \nabla^b h_{ab}^{(T)} = \nabla^b h_{ab}^{(W)}$.  This can be done as follows.
By taking the divergence of Eq.~(\ref{massless}) we find
\beq
(\Box + 3)C_a(h^{(T)}) = 0.  \label{Ca}
\eeq
A complete set of solutions to this equation
is given by $C_a =A_a^{(-6;m\ell\sigma)}$, $m=0,1$, $\ell\geq 1$, and their complex conjugates.  Now, since
\beq
\nabla^b\left(\nabla_{a}A_{b}^{(\mu^2;m\ell\sigma)}+ \nabla_{b}A_{a}^{(\mu^2;m\ell\sigma)}\right) =
\alpha M^2 A_{a}^{(\mu^2;m\ell\sigma)}
\eeq
and that $H_{ab}^{(M^2;m\ell\sigma)}$ are divergence-free, we find
\beq
\nabla^b E^{(2;m\ell\sigma)}_{ab} = \alpha A_a^{(-6;m\ell\sigma)}.
\eeq
Hence, if $C_a(h^{(T)}) = \nabla^b h_{ab}^{(T)} = A_a^{(-6;m\ell\sigma)}$,
then we have $\nabla^b h_{ab}^{(W)}=C_a(h^{(T)})$ by setting
$h_{ab}^{(W)} = \alpha^{-1}E^{(2;m\ell\sigma)}_{ab}$.  It is clear that a similar construction works if
$h_{ab}^{(T)}$ is any linear combination of $A_a^{(-6;m\ell\sigma)}$ and their complex conjugates.

Thus, we have constructed a complete set of solutions to Eq.~(\ref{E-L}), and these solutions are given by
Eqs.~(\ref{scalarsol1}), (\ref{scalarsol2}), (\ref{tensor1}) and (\ref{tensor2}), and their complex conjugates.

\section{The two-point function in the covariant gauge} \label{innerproduct}

In this section we compute the Wightman two-point function for the quantized linearized-gravity field $h_{ab}$ and show that it is physically equivalent to the physical two-point function of Ref.~\cite{HiguchiWeeks} in linearized gravity.

We define the momentum current conjugate to the field $h_{ab}$ by
\begin{eqnarray}
p^{abc} & \equiv & \frac{1}{\sqrt{-g}}\frac{\partial {\cal L}}{\partial (\nabla_a h_{bc})}\nonumber \\
& = & p^{abc}_{\rm inv} + p^{abc}_{\rm gf}, \label{conjugate}
\end{eqnarray}
where
\beqa
p^{abc}_{\rm inv} &  = & - \frac{1}{2}\nabla^a h^{bc} + \frac{1}{2}\left(g^{ab}\nabla_d h^{dc} + g^{ac}\nabla_d h^{db} - g^{bc}\nabla_d h^{da}\right)\nonumber \\
&& - \frac{1}{4}\left( g^{ab}\nabla^c h + g^{ac}\nabla^b h\right) + \frac{1}{2}g^{bc}\nabla^a h,
\label{pinv}  \\
p^{abc}_{\rm gf} & = & - \frac{1}{2\alpha}g^{ab}\left(\nabla_d h^{dc} - \frac{1+\beta}{\beta}\nabla^c h\right)
- \frac{1}{2\alpha}g^{ac}\left(\nabla_dh^{db} - \frac{1+\beta}{\beta}\nabla^b h\right) \nonumber \\
&& + \frac{1+\beta}{\alpha\beta}g^{bc}\left( \nabla_d h^{da} - \frac{1+\beta}{\beta}\nabla^a h\right).
\label{pgf}
\eeqa
Then the Euler-Lagrange equation (\ref{E-L}) can be written as
\beq
\nabla_a p^{abc} - \frac{1}{\sqrt{-g}}\frac{\partial {\cal L}}{\partial h_{bc}}
= 0.
\eeq
The equal-time commutation relations on a $t=$ const Cauchy surface are then given by
\beqa
\left[ h_{ab}(t,\mathbf{x}),h_{cd}(t,\mathbf{x}')\right]
& = & \left[ p^{0ab}(t,\mathbf{x}),p^{0cd}(t,\mathbf{x}')\right] = 0, \label{equaltime1}\\
\left[ h_{ab}(t,\mathbf{x}),p^{0cd}(t,\mathbf{x}')\right]
& = & \frac{\sqrt{-g(x)}}{2}\left(\delta_a^c\delta_b^d + \delta_a^d\delta_b^c\right)\delta(\mathbf{x},\mathbf{x}'),
\label{equaltime2}
\eeqa
where $\delta(\mathbf{x},\mathbf{x}')$ is defined by
\beq
\int_{S^3} d^3\mathbf{x}\,f(\mathbf{x})\delta(\mathbf{x},\mathbf{x}') = f(\mathbf{x}').
\eeq
Here, the $d^3\mathbf{x}$ is the coordinate volume element.  That is, if $\theta_1$, $\theta_2$ and $\theta_3$ are the coordinates on $S^3$, then $d^3\mathbf{x} = d\theta_1d\theta_2d\theta_3$.

For any two solutions $h^{(1)}_{ab}$ and $h^{(2)}_{ab}$ to the Euler-Lagrange
equation (\ref{E-L}) we define the symplectic product by
\beq
(h^{(1)},h^{(2)})_{\rm symp}
= -i \int_\Sigma d\Sigma_a \left( \overline{h^{(1)}_{bc}}p^{(2)abc} - \overline{p^{(1)abc}}h_{bc}^{(2)}\right),
\label{sympprod}
\eeq
where
\beq
p^{(1)abc} \equiv \left. \frac{\partial\mathcal{L}}{\partial(\nabla_a h_{bc})}\right|_{h_{ab}=h^{(1)}_{ab}}
\eeq
on a Cauchy surface $\Sigma$, and similarly for $p^{(2)abc}$.
This symplectic product is independent of the Cauchy surface $\Sigma$ because
the current $\overline{h^{(1)}_{bc}}p^{(2)abc} - \overline{p^{(1)abc}}h_{bc}^{(2)}$ is
conserved~\cite{Friedman,WaldZoupas}.
If $h^{(1)}_{ab}$ and $h^{(2)}_{ab}$ are transverse-traceless `positive-frequency' solutions from Sec.~\ref{physical}, then this symplectic product agrees with Eq.~(\ref{symp1}).

Now, we can expand the quantum field $h_{ab}$ using the solutions found in Sec.~\ref{solutions-covariant}
as follows:
\beqa
h_{ab}(x) & = & \sum_{\ell=2}^\infty\sum_\sigma a^{(TT)}_{\ell\sigma} H_{ab}^{(0;2\ell\sigma)}(x)\nonumber\\
&& + \sum_{m=0}^1\left[\sum_{\ell=2}^\infty \sum_\sigma a^{(T1)}_{m\ell\sigma}E_{ab}^{(1;m\ell\sigma)}(x)
+ \sum_{\ell=1}^\infty \sum_\sigma a^{(T2)}_{m\ell\sigma}E_{ab}^{(2;m\ell\sigma)}(x)\right] \nonumber \\
&& + \sum_{\ell=1}^\infty \sum_\sigma \left[a^{(S1)}_{\ell\sigma}S^{(1\ell\sigma)}_{ab}(x)
+  a^{(S2)}_{\ell\sigma}S^{(2\ell\sigma)}_{ab}(x)\right] + {\rm h.c.}
\eeqa
Let us denote the symplectic product among these solutions as follows:
\beqa
M^{(G;m)}_{AB} & = & (E^{(A;m\ell\sigma)},E^{(B;m\ell\sigma)})_{\rm symp},\,\,\,A,B=1,2,\,\,\,\,m=0,1,\,\,\,
\ell\geq 2, \label{sp1}\\
M^{(G1;m)} & = & (E^{(2;m,\ell=1,\sigma)},E^{(2;m,\ell=1,\sigma)})_{\rm symp},\,\,\,m=0,1,\\
M^{(S)}_{AB} & = & (S^{(A;\ell\sigma)},S^{(B;\ell\sigma)})_{\rm symp},\,\,\,A,B=1,2.
\label{sp2}
\eeqa
[It turns out that these matrix elements are independent of $\ell$ and $\sigma$.
We have already seen that $(H^{(0;2\ell\sigma)},H^{(0;2\ell'\sigma')})_{\rm symp} = \delta^{\ell\ell'}\delta^{\sigma\sigma'}$.]  In Appendix~\ref{ortho-appendix} we show that
$S^{(A;\ell\sigma)}_{ab}$ are orthogonal to the solutions $E^{(A;m\ell\sigma)}_{ab}$ with respect to
the symplectic product (\ref{sympprod}).
Then it is not difficult to show that the equal-time commutation relations (\ref{equaltime1}) and (\ref{equaltime2}) imply
\beqa
\left[a^{(TT)}_{\ell\sigma},a^{(TT)\dagger}_{\ell'\sigma'}\right] & = &
\delta_{\ell\ell'}\delta_{\sigma\sigma'},\\
\left[a^{(T2)}_{m1\sigma},a^{(T2)\dagger}_{m'1\sigma'}\right] & = &
 (M^{(G1;m)})^{-1}\delta_{mm'}\delta_{\sigma\sigma'},\\
\left[a^{(TA)}_{m\ell\sigma},a^{(TB)\dagger}_{m'\ell'\sigma'}\right]
& = & (M^{(G;m)})^{-1}_{AB}\delta_{mm'}\delta_{\ell\ell'}\delta_{\sigma\sigma'},\\
\left[a^{(SA)}_{\ell\sigma},a^{(SB)\dagger}_{\ell'\sigma'}\right]
& = & (M^{(S)})^{-1}_{AB} \delta_{\ell\ell'}\delta_{\sigma\sigma'}.
\eeqa
(See, e.g.\ Ref.~\cite{Higuchi1989}.)
Then the Wightman two-point function for the Bunch-Davies vacuum
can be given as follows:
\beqa
\langle 0|h_{ab}(x)h_{a'b'}(x')|0\rangle
& = & \Delta_{aba'b'}(x,x') \nonumber \\
& = & \Delta^{(\rm phys)}_{aba'b'}(x,x') + \Delta^{(G)}_{aba'b'}(x,x') + \Delta^{(S)}_{aba'b'}(x,x'),
\label{delta-decompose}
\eeqa
where
$\Delta^{(\rm phys)}_{aba'b'}(x,x')$ is the physical two-point function discussed in Sec.~\ref{physical} and where
\beqa
\Delta^{(G)}_{aba'b'}(x,x') & = & \sum_{m=0}^1 \sum_{\sigma}(M^{(G1;m)})^{-1}E^{(2;m,\ell=1,\sigma)}_{ab}(x)
\overline{E^{(2;m,\ell=1,\sigma)}_{a'b'}(x')}\nonumber \\
&& + \sum_{m=0}^1\sum_{\ell=2}^\infty \sum_{\sigma}(M^{(G;m)})^{-1}_{AB}
E^{(A;m\ell\sigma)}_{ab}(x)\overline{E^{(B;m\ell\sigma)}_{a'b'}(x')}, \label{DeltaG}\\
\Delta^{(S)}_{aba'b'}(x,x') & = & \sum_{\ell=0}^\infty \sum_{\sigma}(M^{(S)})^{-1}_{AB}
S^{(A;\ell\sigma)}_{ab}(x)\overline{S^{(B;\ell\sigma)}_{a'b'}(x')}. \label{DeltaS}
\eeqa
Here, the summation over $A$ and $B$ is understood.
Thus, all we need to do is find the matrix elements of the
symplectic product defined by Eqs.~(\ref{sp1})-(\ref{sp2}).

First, we compute  $M^{(G;m)}_{AB}$ for
$\ell\geq 2$ and $M^{(G1;m)}$ and find $\Delta^{(G)}_{aba'b'}(x,x')$ defined by Eq.~(\ref{DeltaG}).
Let us define the invariant and gauge-fixing parts of the symplectic product as follows:
\beqa
(h^{(1)},h^{(2)})_{\rm inv} & = & -i\int_\Sigma d\Sigma_a (\overline{h^{(1)}_{bc}}p^{(2)abc}_{\rm inv} - \overline{p^{(1)abc}_{\rm inv}}h^{(2)}_{bc}),\\
(h^{(1)},h^{(2)})_{\rm gf} & = & -i\int_\Sigma d\Sigma_a (\overline{h^{(1)}_{bc}}p^{(2)abc}_{\rm gf} - \overline{p^{(1)abc}_{\rm gf}}h^{(2)}_{bc}).  \label{inv-inner}
\eeqa
It is well known that if $h^{(k)}_{ab} = \nabla_a A_b^{(k)} + \nabla_bA_a^{(k)}$ for some $A_a^{(k)}$, $k=1,2$, then
$(h^{(1)},h^{(2)})_{\rm inv} = 0$ (see, e.g.\ Ref.~\cite{Henri-Poincare}).   Now, the solutions $E^{(1;m\ell\sigma)}_{ab}$ are
of this form for $m=0,1$ and are divergence-free and traceless.  This implies that $p^{abc}_{\rm gf} = 0$ for these solutions and hence
\beq
M^{(G;m)}_{11} = (E^{(1;m\ell\sigma)},E^{(1;m\ell\sigma)})_{\rm inv} = 0,\,\,\,\ell\geq 2.
\eeq
Next we examine $M^{(G;m)}_{22}$.
We write Eq.~(\ref{tensor2}) as
\beq
E_{ab}^{(2;m\ell\sigma)} = \lim_{M\to 0}\frac{1}{M^2}\left(K^{(M^2; m\ell\sigma)}_{ab}
- H^{(M^2; m\ell\sigma)}_{ab}\right), \label{Kequation}
\eeq
where
\beq
K^{(M^2;m\ell\sigma)}_{ab}  =  \nabla_a A_b^{(\mu^2;m\ell\sigma)} + \nabla_b A_a^{(\mu^2;m\ell\sigma)},
\eeq
with $\mu^2 = \alpha M^2 - 6$ [see Eq.~(\ref{muM})].
We have
\beq
(H^{(M^2;m\ell\sigma)},H^{(M^2;m'\ell'\sigma')})_{\rm symp} = (-1)^{m+1}M^2\delta_{mm'}
\delta_{\ell\ell'}\delta_{\sigma\sigma'} \label{hsymp}
\eeq
for $m,m'=0,1$ from Eq.~(\ref{H1}) because
the symplectic product (\ref{sympprod}) is half the Klein-Gordon inner product (\ref{KGH})
for these solutions.
The symplectic product for the modes $K^{(M^2;m\ell\sigma)}_{ab}$
can be found as follows.  First, since these are of pure-gauge form, we have
\beq
(K^{(M^2;m\ell\sigma)},K^{(M^2; m'\ell'\sigma')})_{\rm inv} = 0.
\eeq
Hence
\beq
(K^{(M^2;m\ell\sigma)},K^{(M^2;m'\ell'\sigma')})_{\rm symp}
= (K^{(M^2;m\ell\sigma)},K^{(M^2;m'\ell'\sigma')})_{\rm gf}.
\eeq
For $h_{ab} = K^{(M^2; m\ell\sigma)}_{ab}$
we find
\beq
p^{abc}_{\rm gf} = -\frac{1}{2}M^2(g^{ab}A^{(\mu^2;m\ell\sigma)c} + g^{ac}A^{(\mu^2;m\ell\sigma)b}).
\eeq
Using this equation and the equality
\beq
A^c \nabla_c A^{\prime a} - A^{\prime c}\nabla_c A^a = \nabla_c (A^{\prime a}A^c - A^a A^{\prime c})\,\,\,
{\rm if}\,\,\,\nabla_c A^c = \nabla_c A^{\prime c} = 0,
\eeq
and the generalized Stokes theorem, we obtain
\beqa
(K^{(M^2;m\ell\sigma)},K^{(M^2;m'\ell'\sigma')})_{\rm symp} & = & -M^2\langle A^{(\mu^2;m\ell\sigma)},A^{(\mu^2;m'\ell'\sigma')}\rangle_{\rm KG}\nonumber \\
&  = & (-1)^mM^2\delta_{mm'}\delta_{\ell\ell'}\delta_{\sigma\sigma'}.  \label{Asymp}
\eeqa
Eq.~(\ref{hsymp}) and this equation together with the fact that there are no cross terms~\cite{Higuchi1989},
i.e.\ $(K^{(M^2;m\ell\sigma)},H^{(M^2;m'\ell'\sigma')})=0$, imply
\beq
M^{(G;m)}_{22} = 0.
\eeq
Finally, Eq.~(\ref{Asymp}) and the fact that there are no cross terms lead to
\beq
M^{(G;m)}_{12} = M^{(G;m)}_{21} = (-1)^{m}.
\eeq

The first line of Eq.~(\ref{Asymp}) is valid for $\ell=1$ as well, but in this case
$H_{ab}^{(M^2;m,\ell=1,\sigma)} =0$ in Eq.~(\ref{Kequation}).    Hence we find, noting
Eq.~(\ref{muM}),
\beq
M^{(G1;m)} = (-1)^m \lim_{M\to 0}\frac{\mu^2+6}{M^2} = (-1)^m \alpha .
\eeq
Clearly the inverse of the matrix $M^{(G;m)}_{AB}$ is itself, and
$(M^{(G1;m)})^{-1} = (-1)^m \alpha^{-1}$.  Hence, from Eq.~(\ref{DeltaG}) we find
\beqa
\Delta^{(G)}_{aba'b'}(x,x') & = & \alpha^{-1}\sum_{m=0}^1 \sum_{\sigma}
(-1)^mE^{(2;m,\ell=1,\sigma)}_{ab}(x)\overline{E^{(2;m,\ell=1,\sigma)}_{a'b'}(x')}\nonumber \\
&& + \sum_{m=0}^1\sum_{\ell=2}^\infty \sum_{\sigma}(-1)^m \left[
E^{(1;m\ell\sigma)}_{ab}(x)\overline{E^{(2;m\ell\sigma)}_{a'b'}(x')}
+ E^{(2;m\ell\sigma)}_{ab}(x)\overline{E^{(1;m\ell\sigma)}_{a'b'}(x')}\right].\nonumber \\
\label{Gexpress}
\eeqa

Now we define the vector two-point function with squared mass $\mu^2$ as
\beqa
\Delta^{(V;\mu^2)}_{aa'}(x,x') & \equiv & \sum_{m=0}^1 \sum_{\ell=1}^\infty
\sum_\sigma \langle A^{(\mu^2;m\ell\sigma)},A^{(\mu^2;m\ell\sigma)}\rangle_{\rm KG}^{-1}
A^{(\mu^2;m\ell\sigma)}_a(x)\overline{A_{a'}^{(\mu^2;m\ell\sigma)}(x')} \nonumber \\
& = &  (\mu^2+6)^{-1} \sum_{m=0}^1 \sum_\sigma (-1)^{m+1}
A^{(\mu^2;m,\ell=1,\sigma)}_a(x)\overline{A^{(\mu^2;m,\ell=1,\sigma)}_{a'}(x')} \nonumber \\
&& + \sum_{m=0}^1\sum_{\ell=2}^\infty \sum_{\sigma} (-1)^{m+1}
A^{(\mu^2;m\ell\sigma)}_a(x)\overline{A^{(\mu^2;m\ell\sigma)}_{a'}(x')}.  \label{defGV}
\eeqa
Let us also write
\beq
\Delta^{(V;(1)\mu^2)}_{aa'}(x,x') \equiv \frac{\partial\ }{\partial\mu^2}\Delta^{(V;\mu^2)}_{aa'}(x,x')
\eeq
and define
\beq
U_{aa'b'}(x,x')  \equiv \sum_{m=0}^1 \sum_{\ell=2}^\infty \sum_\sigma
\left.
A_a^{(-6;m\ell\sigma)}(x)\frac{\partial\ }{\partial M^2}\overline{H_{a'b'}^{(M^2;m\ell\sigma)}(x')}
\right|_{M^2=0}.
\eeq
Then, by Eqs.~(\ref{tensor1}), (\ref{tensor2}) and (\ref{non-indep1}) we find
\beqa
\Delta_{aba'b'}^{(G)}(x,x')
& = & -2\alpha\lim_{\mu^2\to -6}\left[ \nabla_{(a}\nabla_{|a'|}\Delta^{(V;(1)\mu^2)}_{b)b'}(x,x')
+ \nabla_{(a}\nabla_{|b'|}\Delta^{(V;(1)\mu^2)}_{b)a'}(x',x)\right] \nonumber \\
& & +  2\nabla_{(a}U_{b)a'b'}(x,x') + 2\nabla_{(a'}\overline{U_{b')ab}(x',x)}. \label{Gform}
\eeqa
We have used
\beq
\lim_{\mu^2\to -6}(\mu^2+6)^{-1}\nabla_{(a}A^{(\mu^2;m,\ell=1,\sigma)}_{b)}
= \left.\frac{\partial\ }{\partial\mu^2}\nabla_{(a}A^{(\mu^2;m,\ell=1,\sigma)}_{b)}\right|_{\mu^2=-6},
\eeq
which is true because $\nabla_{(a}A^{(-6;m,\ell=1,\sigma)}_{b)} = 0$.
Thus, $\Delta_{aba'b'}^{(G)}$ is of pure-gauge form, i.e.\ is equivalent to $0$ in linearized gravity.

Next, we find the matrix $M^{(S)}_{AB}$ for the solutions $S^{(A;\ell\sigma)}_{ab}$, $A=1,2$,
 given by Eqs.~(\ref{scalarsol1}) and (\ref{scalarsol2}) and use it to find
$\Delta^{(S)}_{aba'b'}(x,x')$ defined by Eq.~(\ref{DeltaS}).
We first express the symplectic product of two solutions
\beq
S_{ab}^{(k)} = \nabla_a\nabla_b B^{(k)} + g_{ab}\Psi^{(k)},\,\,\,k=1,2,
\eeq
in terms of the Klein-Gordon inner product (\ref{scalarKG}).  Let us write the conjugate momentum current
for the solutions $S_{ab}^{(k)}$ as
\beq
p^{(k)abc} = p^{(B,k)abc}_{\rm inv} + p^{(\Psi,k)abc}_{\rm inv} + p^{(k)abc}_{\rm gf},
\eeq
where $p^{(B,k)abc}_{\rm inv}$ and $p^{(\Psi,k)abc}_{\rm inv}$ are the contribution of
$\nabla_b\nabla_c B^{(k)}$ and $g_{bc}\Psi^{(k)}$, respectively, to
$p^{(k)abc}_{\rm inv}$ defined by Eq.~(\ref{pinv}).  The tensor $p^{(k)abc}_{\rm gf}$ is defined by Eq.~(\ref{pgf}).
As noted after Eq.~(\ref{inv-inner}), we have
\beq
\int_\Sigma d\Sigma_a\left[\nabla_b\nabla_c\overline{B^{(1)}}p^{(B,2)abc}_{\rm inv} - \overline{p^{(B,1)abc}_{\rm inv}}\nabla_b\nabla_c
B^{(2)}\right] = 0.
\eeq
Hence,
\beqa
(S^{(1)},S^{(2)})_{\rm symp} & = &
-i \int_\Sigma d\Sigma_a \left[
\nabla_b\nabla_c \overline{B^{(1)}}(p^{(\Psi,2)abc}_{\rm inv} + p^{(2)abc}_{\rm gf})
- (\overline{p^{(\Psi,1)abc}_{\rm inv}}+\overline{p^{(1)abc}_{\rm gf}})\nabla_b\nabla_c B^{(2)}
\right]\nonumber \\
&& - i\int_\Sigma d\Sigma_a \left[
g_{bc}\overline{\Psi^{(1)}}p^{(2)abc} - \overline{p^{(1)abc}}g_{bc}\Psi^{(2)}\right].
\label{S-symp}
\eeqa
By straightforward calculations we find
\beqa
p^{(\Psi,k)abc}_{\rm inv} + p^{(k)abc}_{\rm gf} & = &
- \frac{1}{\beta}g^{bc}\nabla^a\Psi^{(k)},\label{nice2} \\
g_{bc}p^{(k)abc} &  = & - \frac{4}{\beta}\nabla^a\Psi^{(k)} - 3\nabla^a B^{(k)}. \label{nice1}
\eeqa

By substituting these equations into
Eq.~(\ref{S-symp}) and using the field equation (\ref{equation1}) satisfied
by $B^{(1)}$ and $B^{(2)}$, we find
\beq
(S^{(1)},S^{(2)})_{\rm symp} =
3\left(\langle B^{(1)},\Psi^{(2)}\rangle_{\rm KG} + \langle\Psi^{(1)},B^{(2)}\rangle_{\rm KG}\right)
+ (\alpha -3)\langle \Psi^{(1)},\Psi^{(2)}\rangle_{\rm KG}.
\eeq
Note that $\langle B^{(1)},\Psi^{(2)}\rangle_{\rm KG}$, $\langle \Psi^{(1)},B^{(2)}\rangle_{\rm KG}$ and
$\langle \Psi^{(1)},\Psi^{(2)}\rangle_{\rm KG}$ are not conserved individually
though $(S^{(1)},S^{(2)})_{\rm symp}$ is.
The symplectic product for the solutions $S_{ab}^{(A;\ell\sigma)}$, $A=1,2$, given by
Eqs.~(\ref{scalar1})-(\ref{scalar2}) is then
\beq
M^{(S)}_{11}  = 0,\,\,\,
M^{(S)}_{12}  =  3,\,\,\,
M^{(S)}_{22} = \alpha -3.
\eeq
We have used
\beq
\langle \phi^{(\mu^2;\ell\sigma)},\tfrac{\partial\ }{\partial \mu^2}\phi^{(\mu^2;\ell\sigma)}\rangle_{\rm KG}
+ \langle \tfrac{\partial\ }{\partial \mu^2}\phi^{(\mu^2;\ell\sigma)},\phi^{(\mu^2;\ell\sigma)}\rangle_{\rm KG}
= \tfrac{\partial\ }{\partial \mu^2}\langle \phi^{(\mu^2;\ell\sigma)},\phi^{(\mu^2;\ell\sigma)}\rangle_{\rm KG}
= 0
\eeq
in computing $M^{(S)}_{22}$.  The inverse of the matrix $M^{(S)}_{AB}$ is given by
\beq
(M^{(S)})^{-1}_{11} = \tfrac{1}{9}(3 -\alpha),\,\,\,\
(M^{(S)})_{12}^{-1} =  \tfrac{1}{3},\,\,\,
(M^{(S)})_{22}^{-1} = 0.  \label{inverse}
\eeq
Hence, defining the two-point function for the scalar field with mass $\mu$ and its $\mu^2$-derivative as
\beqa
\Delta_{\mu^2}(x,x') & = & \sum_{\ell=0}^\infty \sum_{\sigma} \phi^{(\mu^2;\ell\sigma)}(x)\overline{\phi^{(\mu^2;\ell\sigma)}(x')},\\
\Delta_{\mu^2}^{(1)}(x,x') & = & -\frac{\partial\ }{\partial \mu^2}\Delta_{\mu^2}(x,x'),
\eeqa
and substituting Eq.~(\ref{inverse}) into Eq.~(\ref{DeltaS}), we find
\beqa
\Delta^{(S)}_{aba'b'}(x,x') & = & \nabla_a\nabla_b\nabla_{a'}\nabla_{b'}
\left\{ \frac{3-\alpha}{9}\Delta_{3\beta}(x,x') +\frac{1}{3}\left[4-(\alpha-3)\beta\right]\Delta_{3\beta}^{(1)}(x,x')\right\} \nonumber \\
&& + \frac{1}{3} \left(g_{ab}\nabla_{a'}\nabla_{b'} + g_{a'b'}\nabla_a\nabla_b\right)\Delta_{3\beta}(x,x').
\label{DeltaSresult}
\eeqa
This is clearly of pure-gauge form and generalizes the result obtained in Ref.~\cite{HiguchiKouris2},
where $\Delta^{(S)}_{aba'b'}(x,x')$ for the cases with $(\alpha-3)\beta = 4$ was found.

Thus, we have shown that the two-point function $\Delta_{aba'b'}(x,x')$ in the covariant gauge
given by Eq.~(\ref{delta-decompose}) is equivalent to $\Delta^{({\rm phys})}_{aba'b'}(x,x')$ in linearized gravity because $\Delta^{(G)}_{aba'b'}(x,x')$ and $\Delta^{(S)}_{aba'b'}(x,x')$ given by Eqs.~(\ref{Gform})
and (\ref{DeltaSresult}) are of pure-gauge form.  Notice that the $\alpha\to 0$ limit of
$\Delta_{aba'b'}(x,x')$ is well defined and de~Sitter covariant unless $3\beta = -k(k+3)$, where $k$ is a non-negative integer.
(In the $\alpha\to 0$ limit the gauge condition $\nabla^b h_{ab} - \frac{1+\beta}{\beta}\nabla_a h = 0$ is strictly
enforced on $h_{ab}$.)  Thus, we disagree with the authors of Ref.~\cite{MTW}, who claim that de~Sitter invariance is
broken in the case $\alpha=0$ and $\beta = -2$.

One of the main observations in Ref.~\cite{MTW} is that the scalar two-point function $\Delta_{\mu^2}(x,x')$, which appears in the scalar part $\Delta^{(s)}_{aba'b'}$ of the graviton two-point function, is IR divergent for all negative $\mu^2$.  This is
true if this two-point function is constructed in the Poincar\'e patch of de~Sitter spacetime in momentum expansion.  However, no IR divergences are encountered in the mode-sum construction of $\Delta_{\mu^2}(x,x')$ in global de~Sitter spacetime as shown in Appendix~\ref{app:MTWrebuttal}. (We also show in this appendix that the IR-finite two-point function is recovered even in the Poincar\'e patch by appropriate subtraction.)

Finally, we write $\Delta^{({\rm phys})}_{aba'b'}+ \Delta^{(G)}_{aba'b'}$ in a covariant form.
We first define $\Delta^{(TT;M^2)}_{aba'b'}(x,x')$ to be twice the two-point function for the transverse-traceless
symmetric tensor field with mass $M\neq 0$ satisfying,
\beq
\left[\Box_x - (2+M^2)\right]\Delta^{(TT;M^2)}_{aba'b'}(x,x') = 0.
\eeq
It can be given in the mode-sum form as
\beqa
\Delta^{(TT;M^2)}_{aba'b'}(x,x') & = & 2 \sum_{m=0}^2 \sum_{\ell=2}^\infty
\sum_\sigma \langle H^{(M^2;m\ell\sigma)},H^{(M^2;m\ell\sigma)}\rangle_{\rm KG}^{-1}
H^{(M^2;m\ell\sigma)}_{ab}(x)\overline{H_{a'b'}^{(M^2;m\ell\sigma)}(x')}\nonumber \\
& = & \sum_{\ell=2}^\infty\sum_\sigma
H_{ab}^{(M^2;2\ell\sigma)}(x)\overline{H_{a'b'}^{(M^2;2\ell\sigma)}(x')} \nonumber \\
&& + \frac{1}{M^2}\sum_{m=0}^1\sum_{\ell=2}^\infty \sum_\sigma (-1)^{m+1} H_{ab}^{(M^2;m\ell\sigma)}(x)\overline{H_{a'b'}^{(M^2;m\ell\sigma)}(x')}. \label{defGT}
\eeqa
(See Ref.~\cite{Kogan:2000uy} for an explicit form of $\Delta^{(TT,M^2)}_{aba'b'}$.)
Then we find from Eq.~(\ref{Gexpress}) and the definition $E^{(1;m\ell\sigma)}_{ab} = H^{(0;m\ell\sigma)}_{ab}$ for $m=0,1$ [see Eq.~(\ref{tensor1})]
\beq
\Delta^{({\rm phys})}_{aba'b'}(x,x')
+ \Delta^{(G)}_{aba'b'}(x,x')
= \Delta^{(TT)}_{aba'b'}(x,x') + \Delta^{(V)}_{aba'b'}(x,x'),
\eeq
where
\beqa
\Delta^{(TT)}_{aba'b'}(x,x') & = & \lim_{M^2\to 0}\frac{\partial\ }{\partial M^2}
\left[ M^2 \Delta^{(TT;M^2)}_{aba'b'}(x,x')\right], \label{DeltaTT}\\
\Delta^{(V)}_{aba'b'}(x,x') & = & -2\alpha\lim_{\mu^2\to -6}\left[ \nabla_{(a}\nabla_{|a'|}\Delta^{(V;(1)\mu^2)}_{b)b'}(x,x')
+ \nabla_{(a}\nabla_{|b'|}\Delta^{(V;(1)\mu^2)}_{b)a'}(x,x')\right].
 \label{DeltaV}
\eeqa
These expressions
will be used in Appendix~\ref{consistency} to compare the two-point function found here and the
corresponding result in the Euclidean approach~\cite{HKcovariant}.

\section{Summary} \label{summary}

In this paper we investigated the relationship between the covariant graviton Wightman two-point function and the physical transverse-traceless and synchronous one in global coordinates.  We defined two Wightman graviton two-point functions, $\Delta^{(1)}_{aba'b'}(x,x')$ and $\Delta^{(2)}_{aba'b'}(x,x')$, to be physically equivalent in linearized gravity if they differ by a bi-tensor of the form
$\nabla_{(a}Q_{b)a'b'}(x,x') + \nabla_{(a'}Q_{|ab|b')}(x,x')$ and showed that the covariant two-point function is physically equivalent to the physical two-point function in global coordinates.
Our result is perhaps not surprising, but since there has been much controversy over infrared properties of graviton two-point functions, we believe that it is a worthwhile addition to the body of knowledge about gravitational perturbation in de~Sitter spacetime.

Although our result holds for all $\alpha$ and $\beta$ except $\beta=-L_0(L_0+3)$, $L_0$ nonnegative integer, in global de~Sitter spacetime, one encounters some difficulties if $\beta < 0$ in the Poincar\'e patch because the scalar sector $\Delta_{aba'b'}^{(S)}(x,x')$ becomes tachyonic.  This is also the case for the vector sector $\Delta_{aba'b'}^{(V)}(x,x')$ if $\alpha\neq 0$. (See also Ref.~\cite{Miao:2009hb} for related difficulties with $\alpha\neq 0$.)
However, none of the objections raised in Refs.~\cite{MTW,Miao:2009hb}
are relevant with the choices of gauge parameters $\alpha=0$ and $\beta > 0$ and that the de~Sitter-covariant two-point function can be constructed without any ambiguities even in the Poincar\'e patch.  It will be interesting to construct $\Delta^{(TT)}_{aba'b'}(x,x')$ in Eq.~(\ref{DeltaTT}) in the Poincar\'e patch by the mode-sum method because this is the only place where nontrivial IR issues arise with $\beta > 0$ and $\alpha=0$.

\acknowledgments

We thank Chris Fewster, Don Marolf, Vince Moncrief, Emil Mottola and Evgeny Sklyanin for useful correspondence and discussions and
Chris Fewster and Adrian Ottewill for useful comments on an earlier version of this work~\cite{Faizalthesis}.
One of the authors (A.H.) thanks the Astro-Particle Theory and Cosmology Group and the Department of Applied Mathematics at University of Sheffield, where part of this work was carried out, for their hospitality.

\appendix

\section{Explicit solutions to free-field equations} \label{explicit-solutions}

In this appendix we give the solutions to free-field equations discussed in Sec.~\ref{freefields} explicitly, following Ref.~\cite{HiguchiJMP}
.
The scalar solutions are
\beq
\phi^{(\mu^2;\ell \sigma)} = N_{L_0\ell}(\cosh t)^{-1}{\rm P}_{L_0+1}^{-(\ell+1)}(i\sinh t)Y^{(0\ell\sigma)},  \label{scalarsolution}
\eeq
where $N_{L_0\ell}$ is defined by
\beq
N_{L_0\ell} = \frac{1}{\sqrt{2}}\left[\Gamma(\ell-L_0)\Gamma(\ell+L_0+3)\right]^{1/2}.
\eeq
The transverse vector solutions $A^{(\mu^2;m\ell\sigma)}_a$ are
\beqa
A_0^{(\mu^2;1\ell\sigma)} & = & 0, \label{A1}\\
A_i^{(\mu^2;1\ell\sigma)} & = & \tilde{N}_{L_1\ell}{\rm P}_{L_1+1}^{-(\ell+1)}(i\sinh t)Y^{(1\ell\sigma)}_i,
\label{A2}
\eeqa
where
\beq
\tilde{N}_{L_1\ell} = \begin{cases} N_{L_1\ell} & {\rm if}\,\,\,\ell\geq 2,\\
 \sqrt{\mu^2+6}N_{L_1\ell} & {\rm if}\,\,\,\ell = 1,\end{cases}
\eeq
and
\beqa
A_0^{(\mu^2;0\ell\sigma)} & = & \sqrt{\frac{\ell(\ell+2)}{(L_1+1)(L_1+2)}}
\tilde{N}_{L_1\ell}(\cosh t)^{-2}{\rm P}_{L_1+1}^{-(\ell+1)}(i\sinh t)Y^{(0\ell\sigma)},\\
A_i^{(\mu^2;0\ell\sigma)} & = & - \frac{\tilde{N}_{L_1\ell}}{\sqrt{(L_1+1)(L_1+2)\ell(\ell+2)}}
\left(\frac{\partial\ }{\partial t} + \tanh t\right)
{\rm P}_{L_1+1}^{-(\ell+1)}(i\sinh t)\tilde{\nabla}_i Y^{(0\ell\sigma)}.\nonumber \\ \label{A4}
\eeqa
Finally, the transverse-traceless symmetric tensor solutions are
\beqa
H_{0a}^{(M^2;2\ell\sigma)} & = & 0, \label{firsteq}\\
H_{ij}^{(M^2;2\ell\sigma)} & = & \sqrt{2}N_{L_2\ell}\cosh t\,{\rm P}_{L_2+1}^{-(\ell+1)}(i\sinh t)Y^{(2\ell\sigma)}_{ij}, \label{physicalH}
\eeqa
\beqa
H_{00}^{(M^2;1\ell\sigma)} & = & 0,\\
H_{0i}^{(M^2;1\ell\sigma)} & = & -i\sqrt{(\ell-1)(\ell+3)}
N_{L_2\ell}(\cosh t)^{-1}{\rm P}_{L_2+1}^{-(\ell+1)}(i\sinh t)Y^{(1\ell\sigma)}_i, \label{first-i}\\
H_{ij}^{(M^2;1\ell\sigma)} & = & i\frac{N_{L_2\ell}}{\sqrt{(\ell-1)(\ell+3)}}\cosh t
\left(\frac{\partial\ }{\partial t} +2\tanh t\right)
{\rm P}_{L_2+1}^{-(\ell+1)}(i\sinh t) \nonumber \\
&& \times (\tilde{\nabla}_i Y^{(1\ell\sigma)}_j + \tilde{\nabla}_j Y^{(1\ell\sigma)}_i),
\eeqa
and
\beqa
H_{00}^{(M^2;0\ell\sigma)} & = & -i N'_{L_2\ell}(\cosh t)^{-3}{\rm P}^{-(\ell+1)}_{L_2+1}(i\sinh t)Y^{(0\ell\sigma)},\\
H_{0 i}^{(M^2;0\ell\sigma)}
& = & i N'_{L_2\ell}\frac{(\cosh t)^{-1}}{\ell(\ell+2)}\left(\frac{\partial\ }{\partial t} + \tanh t\right)
{\rm P}^{-(\ell+1)}_{L_2+1}(i\sinh t)\tilde{\nabla}_i Y^{(0\ell\sigma)}, \\
H_{ij}^{(M^2;0\ell\sigma)} & = & -i N'_{L_2\ell}\left\{\frac{3}{2(\ell-1)(\ell+3)}
\left[\frac{\cosh t}{\ell(\ell+2)} \left(\frac{\partial\ }{\partial t} + 2\tanh t\right)
 \left(\frac{\partial\ }{\partial t} + \tanh t\right) + \frac{1}{3\cosh t}\right]\right. \nonumber \\
&& \times  {\rm P}_{L_2+1}^{-(\ell+1)}(i\sinh t)
\left[\tilde{\nabla}_i\tilde{\nabla}_j + \frac{\ell(\ell+2)}{3}\tilde{\eta}_{ij}\right]Y^{(0\ell\sigma)}
 \nonumber \\
&& \left.+ \frac{1}{3\cosh t}\tilde{\eta}_{ij}{\rm P}^{-(\ell+1)}_{L_2+1}(i\sinh t)
Y^{(0\ell\sigma)}\right\}, \label{lasteq}
\eeqa
where
\beq
N'_{L_2\ell} = \sqrt{\frac{4(\ell-1)\ell(\ell+2)(\ell+3)}{3(L_2+1)(L_2+2)}}N_{L_2\ell}.
\eeq

To show Eq.~(\ref{non-indep1}) we used
the associated Legendre equation,
\beq
\left[\frac{d^2\ }{dt^2} +\tanh t\frac{d\ }{dt}
+ \frac{(\ell+1)^2}{\cosh^2 t} - (L+1)(L+2)\right]{\rm P}_{L+1}^{-(\ell+1)}(i\sinh t) = 0,
\label{associated}
\eeq
and the lowering and raising differential operators for the
associated Legendre functions,
\beqa
\cosh t\left[\frac{d\ }{dt} - (L+1)\tanh t\right]{\rm P}_{L+1}^{-(\ell+1)}(i\sinh t) & = &
i(L-\ell){\rm P}_{L}^{-(\ell+1)}(i\sinh t), \label{lowering}\\
\cosh t\left[\frac{d\ }{dt} + (L+1)\tanh t\right]{\rm P}_{L}^{-(\ell+1)}(i\sinh t) & = &
-i(L+\ell+2){\rm P}_{L+1}^{-(\ell+1)}(i\sinh t).
\label{raising}
\eeqa

\section{Orthogonality of scalar-type and tensor-vector-type solutions} \label{ortho-appendix}

In this appendix we show that the symplectic product vanishes between a scalar-type solution $h_{ab}^{(S)}=
S^{(A;\ell\sigma)}_{ab}$
given by Eqs.~(\ref{scalarsol1}) and (\ref{scalarsol2}) and any vector-tensor-type solution $h_{ab}$ satisfying $\nabla_a\nabla_b h^{ab} = 0$ and $h={h^c}_c = 0$.  This result implies that the scalar and tensor-vector sectors can be treated separately as we did.

We consider the symplectic product between a scalar-type solution $h^{(S)}_{ab}= \nabla_a\nabla_b B + g_{ab}\Psi$, with $B$ and $\Psi$ satisfying Eqs.~(\ref{equation1}) and (\ref{equation2}), and the complex conjugate of a vector-tensor-type solution $h_{ab}$:
\beq
(\overline{h},h^{(S)})_{\rm symp} = -i\int_\Sigma d\Sigma_a X^a(h,h^{(S)}),
\eeq
where
\beq
X^a(h,h^{(S)}) \equiv h_{bc}p^{(S) abc} - p^{abc}h^{(S)}_{bc},
\eeq
and where $\Sigma$ is a Cauchy surface, e.g.\ a $t=$ constant\ hypersurface.
The conjugate momentum current $p^{abc}$ here is given by Eq.~(\ref{conjugate}) with the conditions
$\nabla_a\nabla_b h^{ab} = 0$ and $h=0$ imposed.  The contribution to $p_{\rm inv}^{abc}$ defined by
Eq.~(\ref{pinv}) from the part $\nabla_a\nabla_b B$ in $h_{ab}^{(S)}$ can be found as
\beqa
p_{\rm inv}^{(B)abc} & = &
- \frac{1}{2}\nabla^a\nabla^b\nabla^cB - \frac{3}{2}g^{bc}\nabla^a B \nonumber \\
& & + \frac{1}{4}\left[g^{ab}\nabla^c(\Box + 6)B + g^{ac}\nabla^b(\Box + 6)B\right].
\eeqa
The conjugate momentum current for the scalar-type solution $h_{ab}^{(S)}$ is
\beq
p^{(S)abc} = p_{\rm inv}^{(B)abc} + p^{(\Psi)abc}_{\rm inv} + p_{\rm gf}^{(S)abc},
\eeq
where $p^{(\Psi)abc}_{\rm inv}$ is the contribution to $p^{(S)abc}_{\rm inv}$ from
$\nabla_a\nabla_b\Psi$. We have
\beq
p^{(\Psi)abc}_{\rm inv} + p_{\rm gf}^{(S)abc} = - \frac{1}{\beta}\nabla^a \Psi
\eeq
[see Eq.~(\ref{nice2})].
Then we find after a tedious but straightforward calculation
\begin{eqnarray}
X^a(h,h^{(S)}) & = & - \frac{1}{2}h_{bc}\nabla^a\nabla^b\nabla^c B + \frac{1}{2}h^{ab}\nabla_b(\Box + 6)B
\nonumber \\
&& + \frac{1}{2}\nabla^a h^{bc}\nabla_b\nabla_c B + \left(\frac{1}{\alpha}-1\right)\nabla_c h^{bc}\nabla^a\nabla_b B \nonumber \\
&& + \left(\frac{1}{2}-\frac{1+\beta}{\alpha\beta}\right)\nabla_b h^{ab}\Box B
+ \frac{(\alpha-3)\beta-4}{\alpha \beta}\nabla_b h^{ab}\Psi.
\end{eqnarray}

To show that $\int_{\Sigma}d\Sigma_a X^a(h,h^{(S)})=0$ we first note that
\beq
X^a(h,h^{(S)}) = Y^a(h,h^{(S)}) + \nabla_b F^{(1)ab},
\eeq
where
\beq
F^{(1)ab}  =  - \frac{1}{2}\left(h^{bc}\nabla^a \nabla_c B - h^{ac}\nabla^b \nabla_c B\right)
+ \frac{1}{2}\left(\nabla^a h^{bc} \nabla_c B - \nabla^b h^{ac}\nabla_c B\right),
\eeq
and, with the definition $C^a = \nabla_b h^{ab}$,
\beqa
Y^a(h,h^{(S)}) & = & \left( \frac{1}{\alpha}-\frac{1}{2}\right)C^b\nabla^a \nabla_b B
+ \left( \frac{1}{2}-\frac{1+\beta}{\alpha\beta}\right)C^a \Box B \nonumber \\
&& + \left[- \frac{1}{2\alpha}\nabla^a C^b + \left( \frac{1}{2} - \frac{1}{2\alpha}\right)\nabla^b C^a\right]
\nabla_b B \nonumber \\
&& - \frac{4- (\alpha-3)\beta}{\alpha\beta}C^a\Psi.
\eeqa
We have used the field equation (\ref{E-L}) to solve for $\Box h_{ab}$.  Since $F^{(1)ab}$ is an anti-symmetric tensor, we have
\beq
\int_\Sigma d\Sigma_a \nabla_b F^{(1)ab} = 0
\eeq
by the generalized Stokes theorem.  Hence
\beq
(\overline{h},h^{(S)})_{\rm symp} =- i \int_\Sigma d\Sigma_a Y^a(h,h^{(S)}).
\label{sympY}
\eeq
Next we find
\beq
Y^a(h,h^{(S)}) = \nabla_b F^{(2)ab} - \frac{1}{\alpha\beta}C^a
\left\{ (\Box - 3\beta)B + \left[4-(\alpha-3)\beta\right]\Psi\right\},  \label{close}
\eeq
where
\beq
F^{(2)ab}  =  \left(\frac{1}{\alpha}-\frac{1}{2}\right)(C^b\nabla^a B - C^a \nabla^b B)
+ \frac{1}{2\alpha}B(\nabla^b C^a - \nabla^a C^b)
\eeq
by using the equation
\beq
\nabla_b(\nabla^b C^a - \nabla^a C^b) = -6 C^a
\eeq
[see Eq.~(\ref{Ca})].  Finally by Eq.~(\ref{equation1}) and anti-symmetry of $F^{(2)ab}$, we find
$(\overline{h},h^{(S)})_{\rm symp} = 0$
from Eqs.~(\ref{sympY}) and (\ref{close}).

\section{Two-point function for tachyonic scalar field} \label{app:MTWrebuttal}

It has been pointed out in Ref.~\cite{MTW} that the two-point function for the scalar field with negative mass squared is IR divergent if it is expanded in terms of momentum eigenfunctions in the Poincar\'e patch and that as a result the de~Sitter invariant graviton two-point function is IR divergent for $\beta < 0$.  This is true even if $\beta$ is not one of the discrete values for which it is IR divergent in the Euclidean approach~\cite{Allen86b,HKcovariant}.

In this appendix we verify that \emph{in global coordinates} the de~Sitter-invariant two-point function is IR finite even if the field is tachyonic unless the mass squared $\mu^2$ is
of the form $-L_0(L_0+3)$, $L_0=0,1,2,\ldots$ by explicitly constructing it.  We also point out that this two-point function is recovered also in the Poincar\'e patch if an appropriate IR subtraction is made.

\subsection{Construction of the scalar two-point function in global coordinates}

We first show that the scalar two-point function can be constructed by the mode-sum method in global coordinates without any IR divergences even with tachyonic mass unless the mass squared satisfies $\mu^2= -L_0(L_0+3)$, $L_0=0,1,2,\ldots$.

We write the metric on the unit $S^3$ as
\begin{equation}
d\Omega^2 = d\chi^2 + \sin^2\chi(d\theta^2 + \sin^2\theta\,d\varphi^2),
\end{equation}
where $0 \leq \chi \leq \pi$ and where $\theta$ and $\varphi$ are the usual spherical polar coordinates on $S^2$.
The `positive-frequency' mode functions corresponding to the Bunch-Davies vacuum are given by
Eq.~(\ref{scalarmodes}):
\begin{equation}
\Phi^{(\ell\ell_2 m)}(t,\chi,\theta,\varphi)
= \frac{1}{\cosh t}{\rm P}_{L_0+1}^{-(\ell+1)}(i\sinh t)Y^{(0\ell\ell_2 m)}(\chi,\theta,\varphi),
\,\,\,L_0 = - \frac{3}{2}+\sqrt{\frac{9}{4}-\mu^2},
\end{equation}
where
\begin{equation}
Y^{(0\ell\ell_2 m)}(\chi,\theta,\chi) = \frac{\ell+1}
{\sqrt{\sin\chi}}{\rm P}_{\ell+\frac{1}{2}}^{-(\ell_2+\frac{1}{2})}
(\cos\chi)Y_{\ell_2 m}(\theta,\varphi).
\end{equation}
The $Y_{\ell_2 m}(\theta,\varphi)$ are the standard spherical harmonics on $S^2$.

The Wightman two-point function with one point at $\chi=0$ is given as~\cite{HiguchiJMP}
\begin{equation}
G(t_1,t_2,\chi) \equiv \sum_{\ell=0}^\infty \frac{\Gamma(\ell-L_0)\Gamma(\ell+L_0+3)}{2}
\Phi^{(\ell 00)}(t_1,\chi,\theta_1,\varphi_1)
\overline{\Phi^{(\ell 00)}(t_2,0,\theta_2,\varphi_2)}.  \label{modesumsum}
\end{equation}
If $L_0>0$, i.e.\ if $\mu^2< 0$,
then some modes have negative coefficients, i.e.\ have negative norm.

We assume that $L_0$ is not an integer. If $L_0$ is an integer, then this two-point function is indeed IR divergent. We note in passing that the modes $\Phi^{(\ell\ell_2 m)}$ with positive norm form a unitary representation of the de~Sitter group if $L_0$ \emph{is} an integer whereas for a positive non-integer value of $L_0$ no unitary representation exists because of the negative norm modes~\cite{HiguchiJMP,Bros:2010wa}.

Since only the $\ell_2=0$ modes contribute in Eq.~(\ref{modesumsum}),
the function $G(t_1,t_2,\chi)$ is independent of $\theta_1$, $\theta_2$, $\varphi_1$ and $\varphi_2$. By noting that
\begin{equation}
{\rm P}_{\ell+\frac{1}{2}}^{-\frac{1}{2}}(\cos\chi) =
 \sqrt{\frac{2}{\pi}}\frac{\sin(\ell+1)\chi}{(\ell+1)\sin\chi},
\end{equation}
we obtain
\begin{eqnarray}
G(t_1,t_2,\chi) & = &
\frac{1}{4\pi^2}\sum_{\ell=0}^\infty (\ell+1)\Gamma(\ell-L_0)\Gamma(\ell+L_0+3) \nonumber \\
&& \times \frac{1}{\cosh t_1}{\rm P}_{L_0+1}^{-(\ell+1)}(i\sinh t_1 + \epsilon)
\frac{1}{\cosh t_2}{\rm P}_{L_0+1}^{-(\ell+1)}(-i\sinh t_2 + \epsilon) \frac{\sin(\ell+1)\chi}{\sin\chi},
\nonumber \\ \label{first-go}
\end{eqnarray}
where we inserted the `infinitesimal' positive number $\epsilon$ for UV regularization. This series can be
shown to be convergent by using
\begin{eqnarray}
{\rm P}_{L_0+1}^{-(\ell+1)}(z) & = & \frac{1}{(\ell+1)!}\left( \frac{1 - z}{1+z}\right)^{\ell+1}
F(-L_0-1,L_0+2; \ell+2;(1-z)/2) \nonumber \\
& \approx & \frac{1}{(\ell+1)!}
\left( \frac{1 - z}{1+z}\right)^{\ell+1}\,\,\,{\rm if}\,\,\ell\gg 1.
\label{LeGendre}
\end{eqnarray}
By using the identity $\Gamma(u)\Gamma(1-u)=\pi/\sin \pi u$,
we find that Eq.~(\ref{first-go}) can be written
\begin{eqnarray}
G(t_1,t_2,\chi) & = &
-\frac{\Gamma(-L_0-1)\Gamma(L_0+2)}{4\pi^2\cosh t_1\cosh t_2\sin\chi}
\sum_{\ell=0}^\infty (\ell+1)\frac{\Gamma(L_0+\ell+3)}{\Gamma(L_0-\ell+1)} \nonumber \\
&& \,\,\,\times{\rm P}_{L_0+1}^{-(\ell+1)}(i\sinh t_1 + \epsilon)
{\rm P}_{L_0+1}^{-(\ell+1)}(-i\sinh t_2 + \epsilon)\sin[(\ell+1)(\pi-\chi)]. \label{second-go}
\end{eqnarray}
Now, an addition theorem for the associated Legendre functions (8.794.1 of Ref.~\cite{GR}) can be adapted to the series here as
\begin{eqnarray}
&& {\rm P}_{L_0+1}(\sinh t_1 \sinh t_2 - \cosh t_1\cosh t_2 \cos\chi + i\epsilon(t_1-t_2)) \nonumber \\
&& = {\rm P}_{L_0+1}(i\sinh t_1 + \epsilon){\rm P}_{L_0+1}(-i\sinh t_2+\epsilon)
\nonumber\\
&& \,\,\, + 2\sum_{\ell=0}^\infty
\frac{\Gamma(L_0+\ell+3)}{\Gamma(L_0-\ell+1)}
{\rm P}_{L_0+1}^{-(\ell+1)}(i\sinh t_1 + \epsilon)
{\rm P}_{L_0+1}^{-(\ell+1)}(-i\sinh t_2 + \epsilon)\cos[(\ell+1)(\pi - \chi)]. \nonumber \\
\end{eqnarray}
By differentiating both sides with respect to $\chi$ and substituting the result into
Eq.~(\ref{second-go}) we obtain
\begin{eqnarray}
G(t_1,t_2,\chi) & = &
-\frac{\Gamma(-L_0-1)\Gamma(L_0+2)}{8\pi^2\cosh t_1\cosh t_2\sin\chi} \nonumber \\
&& \times \frac{d\ }{d\chi}{\rm P}_{L_0+1}(\sinh t_1 \sinh t_2 - \cosh t_1\cosh t_2 \cos\chi + i\epsilon(t_1-t_2)).
\end{eqnarray}
Finally, by using Eq.~(\ref{LeGendre}) with
${\rm P}_{L_0+1}(z) = {\rm P}_{L_0+1}^0(z)$ in Eq.~(\ref{second-go}) and
\begin{equation}
\frac{d\ }{du}F(a,b;c;u) = \frac{ab}{c}F(a+1,b+1;c+1;u),
\end{equation}
we find
\begin{eqnarray}
G(t_1,t_2,\chi) & = & \frac{\Gamma(-L_0)\Gamma(L_0+3)}{16\pi^2}
F(-L_0,L_0+3;2;( 1+ Z -i\epsilon(t_1-t_2))/2), \label{globalG}\\
Z & \equiv & -\sinh t_1\sinh t_2 + \cosh t_1\cosh t_2\cos\chi,
\end{eqnarray}
which is the standard result~\cite{BunchDavies,AllenJacobson}. Note that our result is valid also for $L_0> 0$, i.e.\ for tachyonic scalar fields, as long as $L_0$ is not an integer.

\subsection{Two-point function for tachyonic scalar field in the
Poinacar\'e patch}

In this subsection we show that, even though the two-point function for tachyonic scalar field is IR divergent in the momentum expansion in the Poincar\'e patch, one can still recover the two-point function found in the previous subsection by subtracting the IR divergent terms.

In the spatially-flat coordinate system the metric of de~Sitter spacetime can be given as
\begin{equation}
ds^2 = \frac{1}{\eta^2}(-d\eta^2 + d\mathbf{x}^2),\,\,\, \eta \in (-\infty,0).
\end{equation}
The Wightman two-point function between points $(\eta_1,\mathbf{x}_1)$ and $(\eta_2,\mathbf{x}_2)$ with
$\|\mathbf{x}_1-\mathbf{x}_2\| = r$ is found as
\begin{equation}
G_{\rm flat}(\eta_1,\eta_2,r)
= \frac{(\eta_1\eta_2)^{3/2}}{8\pi r}\int_0^\infty dk\,k\sin kr H_\nu^{(1)}(-k\eta_1)\overline{H_\nu^{(1)}(-k\eta_2)}, \label{sp_flat_G}
\end{equation}
where
\begin{equation}
\nu \equiv L_0 + \frac{3}{2} = \sqrt{\frac{9}{4}-\mu^2}.
\end{equation}
The Hankel function is given in terms of the Bessel function as
\begin{equation}
H_\nu^{(1)}(u) = \frac{i}{\sin\pi \nu}\left[ e^{-i\pi\nu}J_\nu(u) - J_{-\nu}(u)\right].
\end{equation}
The integral (\ref{sp_flat_G}) converges if $\mu^2 > 0$ and the result of the integral is known to agrees with
$G(t_1,t_2,\chi)$ in Eq.~(\ref{globalG})~\cite{BunchDavies} with
\begin{equation}
Z = \frac{\eta_1^2+\eta_2^2 - r^2}{2\eta_1\eta_2}
\end{equation}
in this case. We have
\begin{equation}
J_{-\nu}(u) \approx \frac{1}{\Gamma(1-\nu)}\left( \frac{2}{u}\right)^{\nu},\,\,\,|u|\ll 1.
\end{equation}
Hence the integral (\ref{sp_flat_G}) diverges in the infrared if $\nu \geq \tfrac{3}{2}$, i.e.\ if $\mu^2\leq 0$.

Let us first separate out the term causing the IR divergences as
\begin{equation}
G_{\rm flat}(\eta_1,\eta_2,r) = G^{({\rm reg})}_{\rm flat}(\eta_1,\eta_2,r)
+ G^{(\infty)}_{\rm flat}(\eta_1,\eta_2,r),
\end{equation}
where the IR-divergent contribution for $\nu \geq \tfrac{3}{2}$ is given by
\begin{equation}
G^{(\infty)}_{\rm flat}(\eta_1,\eta_2,r)
=
\frac{(\eta_1\eta_2)^{3/2}}{8\pi r\sin \pi\nu}\int_0^\lambda dk\,k\sin kr J_{-\nu}(-k\eta_1)
J_{-\nu}(-k\eta_2),\,\,\,\lambda >0.  \label{div-int}
\end{equation}
(The case with integer
$\nu$ needs to be treated as a limit of cases with non-integer $\nu$.)
The function $G^{({\rm reg})}_{\rm flat}(\eta_1,\eta_2,r)$ is the IR-regularized two-point function with the IR cutoff $\lambda$.
If ${\rm Re}\,\nu < \tfrac{3}{2}$, then the integral in Eq.~(\ref{div-int}) is convergent and tends to zero as $\lambda\to 0$.  Now, this can be analytically continued to ${\rm Re}\,\nu > \tfrac{3}{2}$ as
\begin{equation}
G^{({\rm sub})}_{\rm flat}(\eta_1,\eta_2,r)
\equiv \frac{(\eta_1\eta_2)^{3/2}}{8\pi r\sin\pi\nu(1+ e^{-2\pi i\nu})}
\int_C dk\,k\sin kr J_{-\nu}(-k\eta_1)J_{-\nu}(-k\eta_2),
\end{equation}
where $C$ is a path on the complex $k$-plane from $-\lambda$ to $\lambda$ that avoids the origin in the upper
half-plane. This means that the two-point function defined by
\begin{equation}
G_{\rm flat}^{({\rm inv})}(\eta_1,\eta_2,r)
\equiv G^{({\rm reg})}_{\rm flat}(\eta_1,\eta_2,r) + G^{({\rm sub})}_{\rm flat}(\eta_1,\eta_2,r)
\end{equation}
is the two-point function $G(t_1,t_2,\chi)$ given by Eq.~(\ref{globalG}) expressed in spatially-flat coordinates even for ${\rm Re}\,\nu > \tfrac{3}{2}$.
Thus, $G_{\rm flat}^{({\rm reg})}$ is the IR-regularized two-point function
as mentioned before and
$G_{\rm flat}^{({\rm sub})}$ is the IR-subtraction term needed to recover the de~Sitter-invariant two-point function.  Note that this scheme does not work
if $\nu$ is a half-odd-integer because
$G^{({\rm sub})}_{\rm flat}(\eta_1,\eta_2,r)$ is infinite in this case.

Let us examine the IR-subtraction term $G^{({\rm sub})}_{\rm flat}$ more closely for $\tfrac{3}{2} < \nu < \tfrac{5}{2}$ in the limit $\lambda\to 0$. Choosing $C$ to be the upper semicircle from $-\lambda$ to $\lambda$, we find
\begin{equation}
G^{({\rm reg})}_{\rm flat}(\eta_1,\eta_2,r)
 = - \frac{[\Gamma(\nu)]^2\lambda^{3-2\nu}}{\pi^2(2\nu-3)}\left( \frac{4}{\eta_1\eta_2}\right)^{\nu} + O(\lambda^{5-2\nu}).
\end{equation}
Note that $\lambda^{5-2\nu}\to 0$ as $\lambda \to 0$ by our assumption
$\nu < \tfrac{5}{2}$.
Hence we have
\begin{eqnarray}
G_{\rm flat}^{({\rm inv})}(\eta_1,\eta_2,r)
& = & \lim_{\lambda\to 0}\left[\frac{(\eta_1\eta_2)^{3/2}}{8\pi r}\int_\lambda^\infty dk\,k\sin kr H_\nu^{(1)}(-k\eta_1)\overline{H_\nu^{(1)}(-k\eta_2)}\right. \nonumber \\
&& \,\,\,\,\,\,\,\,\,\,\,\,\,\,\,\,\left. - \frac{[\Gamma(\nu)]^2}{\pi^3(2\nu-3)}\left(
\frac{\lambda^2\eta_1\eta_2}{4}\right)^{\tfrac{3}{2}-\nu}
\right].  \label{Ginv}
\end{eqnarray}
Thus, to recover the de~Sitter covariant two-point function for $\frac{3}{2}< \nu < \frac{5}{2}$ we need to remove the IR divergences by subtracting some zero-mode
contribution.

Finally, we verify that the large $r$ behavior of $G_{\rm flat}^{(\rm inv)}(\eta_1,\eta_2,r)$ is correctly reproduced by Eq.~(\ref{Ginv}).  From Eq.~(\ref{globalG}) we find, using a transformation formula for hypergeometric functions and the doubling formula for the Gamma function,
\begin{equation}
G^{({\rm inv})}_{\rm flat}(\eta_1,\eta_2,r)
\approx \frac{1}{4\pi^{5/2}}\Gamma(\tfrac{3}{2}-\nu)
\Gamma(\nu)\left(\frac{r^2}{\eta_1\eta_2}\right)^{\nu - \frac{3}{2}}.  \label{r-dependence}
\end{equation}
By examining the $\eta_1\eta_2$ dependence of this term we find that this term comes entirely from the leading term in the $k$-expansion of $H_\nu^{(1)}(-k\eta_1)\overline{H_\nu^{(1)}(-k\eta_2)}$ in Eq.~(\ref{Ginv}).
Thus, we find
\begin{equation}
G_{\rm flat}^{({\rm inv})}(\eta_1,\eta_2,r)
\approx \lim_{\lambda\to 0}\left\{\frac{[\Gamma(\nu)]^2}{\pi^3}
\left(\frac{\eta_1\eta_2}{4}\right)^{\frac{3}{2}-\nu}
\int_\lambda^\infty dk\,k^{2-2\nu}\frac{\sin kr}{kr}
- \frac{[\Gamma(\nu)]^2}{\pi^3(2\nu-3)}
\left(\frac{\lambda^2\eta_1\eta_2}{4}\right)^{\frac{3}{2}-\nu}\right\}.
\end{equation}
Upon integration by parts the second term cancels out the boundary term, and we obtain
\begin{equation}
G_{\rm flat}^{({\rm inv})}(\eta_1,\eta_2,r)
\approx
\frac{[\Gamma(\nu)]^2}{\pi^3(2\nu-3)}
\left(\frac{\eta_1\eta_2}{4r^2}\right)^{\frac{3}{2}-\nu}
\int_0^\infty du\,u^{2-2\nu}\left( \cos u - \frac{\sin u}{u}\right),  \label{intermed}
\end{equation}
where we have let $u\equiv kr$. We find Eq.~(\ref{r-dependence}) by evaluating this integral.

\section{Comparison with the Euclidean approach} \label{consistency}

In this paper we found the covariant
graviton two-point function using the mode-sum method.  It can be written as
\beq
\Delta_{aba'b'}(x,x') = \Delta^{(TT)}_{aba'b'}(x,x') + \Delta_{aba'b'}^{(V)}(x,x') + \Delta_{aba'b'}^{(S)}(x,x'),  \label{Lorentzian}
\eeq
where  $\Delta^{(TT)}_{aba'b'}$, $\Delta^{(V)}_{aba'b'}$ and $\Delta^{(S)}_{aba'b'}$ are given by
Eqs.~(\ref{DeltaTT}), (\ref{DeltaV}) and (\ref{DeltaSresult}), respectively.
Now, this two-point function can also be found in the Euclidean approach.  In this approach $\Delta_{aba'b'}(x,x')$ can also be given as a sum of three parts:
\beq
\Delta_{aba'b'}(x,x') = G^{(TT)}_{aba'b'}(x,x') + G^{(V)}_{aba'b'}(x,x') + G^{(S)}_{aba'b'}(x,x').
\label{Euclidean}
\eeq
(See, e.g.\ Refs.~\cite{HKcovariant,AllenTuryn}.  Our graviton two-point functions are twice that of Ref.~\cite{AllenTuryn}.)
The function $G^{(TT)}_{aba'b'}(x,x')$ is transverse-traceless and $G^{(V)}_{aba'b'}(x,x')$ is a symmetric derivative in each of the sets of indices $(ab)$ and $(a'b')$ of a
vector two-point function like $\Delta^{(V)}_{aba'b'}(x,x')$ in the mode-sum case.  However, these functions are not equal to
$\Delta_{aba'b'}^{(TT)}(x,x')$ and $\Delta_{aba'b'}^{(V)}(x,x')$, respectively.
We also find that the scalar part in
the Euclidean approach,
$G^{(S)}_{aba'b'}(x,x')$, given in Ref.~\cite{HKcovariant} is different from $\Delta^{(S)}_{aba'b'}(x,x')$.
In this appendix we verify that Eqs.~(\ref{Lorentzian}) and (\ref{Euclidean}) give the same two-point function
for spacelike-separated points $x$ and $x'$ in spite of these differences.

Let us describe the difference between $\Delta^{(S)}_{aba'b'}$ given by Eq.~(\ref{DeltaSresult}) and the scalar part $G^{(S)}_{aba'b'}$ in the Euclidean approach.
For spacelike-separated points $x$ and $x'$, the
two-point function $\Delta_{\mu^2}(x,x')$ for scalar field
in de~Sitter spacetime is identical to the corresponding Green's function on $S^4$ as a function of the geodesic distance between $x$ and $x'$.
If we let $\psi^{(n\nu)}(x)$, $n=0,1,2,\ldots$, be a complete set of orthonormal
scalar modes on $S^4$ satisfying
\beq
\left[\Box + n(n+3)\right]\psi^{(n\nu)}(x) = 0,\,\,\,n=0,1,2,\ldots,
\eeq
where $\nu$ represents all labels other than $n$,
and
\beq
\int_{S^4}dS\, \overline{\psi^{(n\nu)}(x)}\psi^{(n'\nu')}(x) = \delta^{nn'}\delta^{\nu\nu'},
\eeq
then one can readily see that the equation for the Green's function
\beq
(-\Box_x + \mu^2)\Delta_{\mu^2}(x,x') = \delta(x,x'),
\eeq
where
\beq
\delta(x,x') = \sum_{n=0}^\infty\sum_\nu \psi^{(n\nu)}(x)\overline{\psi^{(n\nu)}(x')},
\eeq
is uniquely solved by
\beq
\Delta_{\mu^2}(x,x') = \sum_{n=0}^\infty \sum_\nu \frac{\psi^{(n\nu)}(x)\overline{\psi^{(n\nu)}(x')}}{n(n+3)+\mu^2}.  \label{Euclscalar}
\eeq
We define
\beqa
\Delta_{\mu^2}^{-}(x,x') & \equiv & \sum_{n=1}^\infty \sum_{\nu}\frac{\psi^{(n\nu)}(x)\overline{\psi^{(n\nu)}(x')}}{n(n+3)+\mu^2}, \label{defminus}\\
\Delta_{\mu^2}^{--}(x,x') & \equiv & \sum_{n=2}^\infty \sum_{\nu}\frac{\psi^{(n\nu)}(x)\overline{\psi^{(n\nu)}(x')}}{n(n+3)+\mu^2}. \label{defminusminus}
\eeqa
Then the scalar part in the Euclidean approach, $G^{(S)}_{aba'b'}(x,x')$, is given in Ref.~\cite{HKcovariant}
as
\beqa
G^{(S)}_{aba'b'}(x,x') & = & \Delta^{(S)}_{aba'b'}(x,x')
+ \frac{\alpha}{9}\nabla_a\nabla_b\nabla_{a'}\nabla_{b'}\Delta_0^{-}(x,x')\nonumber \\
&& - \frac{1}{3}\left(\nabla_a\nabla_b - \frac{1}{4}g_{ab}\Box\right)
\left(\nabla_{a'}\nabla_{b'} - \frac{1}{4}g_{a'b'}\Box'\right)\Delta_{-4}^{--}(x,x').
\eeqa
Hence, the Euclidean and mode-sum approaches will be consistent
with each other if
\beqa
G^{(V)}_{aba'b'}(x,x') & = & \Delta^{(V)}_{aba'b'}(x,x')
- \frac{\alpha}{9}\nabla_a\nabla_b\nabla_{a'}\nabla_{b'}\Delta_0^{-}(x,x'),\label{Vconsistent}\\
G^{(TT)}_{aba'b'}(x,x') & = &
\Delta^{(TT)}_{aba'b'}(x,x')
+ \frac{1}{3}\left(\nabla_a\nabla_b - \frac{1}{4}g_{ab}\Box\right)
\left(\nabla_{a'}\nabla_{b'} - \frac{1}{4}g_{a'b'}\Box'\right)\Delta_{-4}^{--}(x,x'). \nonumber \\
\label{TTconsistent}
\eeqa
We will verify these relations in the rest of this appendix.

To show Eq.~(\ref{Vconsistent}) we first need to
define the Green's function $G^{(V;\mu^2)}_{aa'}(x,x')$ for the transverse vector field with mass $\mu^2$
in the Euclidean approach.  Let
$V_a^{(n\nu)}(x)$, $n=1,2,\ldots$, form a complete orthonormal
set of transverse solutions to the eigenvalue equation
on $S^4$,
\beq
\nabla^b(\nabla_a V_b^{(n\nu)} - \nabla_b V_a^{(n\nu)}) = (n+1)(n+2)V_a^{(n\nu)},\,\,\,n=1,2,\ldots,
\eeq
satisfying $\nabla^a V_a^{(n\nu)} = 0$ and
\beq
\int_{S^4} dS\, \overline{V_a^{(n\nu)}(x)}V^{(n'\nu')a}(x) = \delta^{nn'}\delta^{\nu\nu'}.
\eeq
Then, we define the transverse Green's function for the operator
\beq
L^{(V)b}_a V_b \equiv \nabla^b(\nabla_a V_b - \nabla_b V_a) + \mu^2 V_a
\eeq
by
\beq
G^{(V;\mu^2)}_{aa'}(x,x') \equiv \sum_{n=1}^\infty \sum_\nu
\frac{V_a^{(n\nu)}(x)\overline{V_{a'}^{(n\nu)}(x')}}{(n+1)(n+2)+\mu^2}. \label{VecG}
\eeq
This Green's function satisfies
\beq
L^{(V)b}_a G^{(V;\mu^2)}_{ba'}(x,x') = \delta^{(V)}_{aa'}(x,x'),
\eeq
where
\beq
\delta^{(V)}_{aa'}(x,x') = \sum_{n=1}^\infty \sum_\nu V_a^{(n\nu)}(x)\overline{V_{a'}^{(n\nu)}(x')}.
\eeq
On the other hand, the Euclidean Green's function $\Delta^{(V;\mu^2)}_{aa'}(x,x')$ that becomes the Feynman
propagator and hence the Wightman two-point function for spacelike-separate points after appropriate analytic continuation satisfies~\cite{AllenTuryn}
\beq
L^{(V)b}_a \Delta^{(V;\mu^2)}_{ba'}(x,x') = \delta_{aa'}(x,x'),
\eeq
where
\beqa
\delta_{aa'}(x,x') & = & \delta^{(V)}_{aa'}(x,x') + \sum_{n=1}^\infty \sum_\nu
\frac{\nabla_a\psi^{(n\nu)}(x)\nabla_{a'}\overline{\psi^{(n\nu)}(x')}}{n(n+3)} \nonumber \\
& = & \delta^{(V)}_{aa'}(x,x') + \nabla_a \nabla_{a'}\Delta_0^{-}(x,x').
\eeqa
The two-point function $\Delta_0^{-}(x,x')$ is defined by Eq.~(\ref{defminus}).  By noting that
\beq
L^{(V)b}_a \nabla_b\nabla_{a'}\Delta_0^{-}(x,x') = \mu^2\nabla_a\nabla_{a'}\Delta_0^{-}(x,x'),
\eeq
we readily find~\cite{AllenTuryn}
\beq
G^{(V;\mu^2)}_{aa'}(x,x') = \Delta^{(V;\mu^2)}_{aa'}(x,x') - \frac{1}{\mu^2}\nabla_a\nabla_{a'}\Delta_0^{-}(x,x').
\label{Vdifference}
\eeq
The vector part of the propagator in the Euclidean approach is~\cite{HKcovariant,AllenTuryn}
\beq
G^{(V)}_{aba'b'}(x,x') =
4 \alpha \sum_{n=2}^\infty \sum_\nu
\frac{\nabla_{(a}V_{b)}^{(n\nu)}(x)\nabla_{(a'}\overline{V_{b')}^{(n\nu)}(x')}}
{\left[(n+1)(n+2)-6\right]^2}.
\eeq
Note that there is no contribution from the vectors $V_a^{(n=1,\nu)}$ because they are Killing vectors
on $S^4$.  Using the definition (\ref{VecG}), we find
\beq
G^{(V)}_{aba'b'}(x,x') = - 2\alpha \lim_{\mu^2\to -6}\frac{\partial\ }{\partial \mu^2}
\left[\nabla_{(a}\nabla_{|a'|} G^{(V;\mu^2)}_{b)b'}(x,x')
+ \nabla_{(a}\nabla_{|b'|}G^{(V;\mu^2)}_{b)a'}(x,x')\right].
\eeq
[Notice the similarity of this equation with Eq.~(\ref{DeltaV}).]  From Eq.~(\ref{Vdifference}) we
readily find Eq.~(\ref{Vconsistent}).

Next we show Eq.~(\ref{TTconsistent}).  The transverse-traceless part of the two-point function in the
Euclidean approach is~\cite{AllenTuryn}
\beq
G^{(TT)}_{aba'b'}(x,x')  = 2\sum_{n=2}^\infty \sum_\nu
\frac{K_{ab}^{(n\nu)}(x)\overline{K_{a'b'}^{(n\nu)}(x')}}{n(n+3)},
\eeq
where $K_{ab}^{(n\nu)}(x)$ form a complete orthonormal
set of transverse-traceless eigenfunctions satisfying
\beqa
L_{ab}^{({\rm inv}) cd}K_{cd}^{(n\nu)}
& = & (-\Box + 2)K_{ab}^{(n\nu)} \nonumber \\
& = & n(n+3)K_{ab}^{(n\nu)},
\eeqa
and
\beq
\int_{S^4}dS\,\overline{K_{ab}^{(n\nu)}(x)}K^{(n'\nu')ab}(x) = \delta^{nn'}\delta^{\nu\nu'}.
\eeq
It is convenient to define the massive Green's function $G^{(TT;M^2)}_{aba'b'}(x,x')$ by
\beqa
L_{ab}^{(M^2)cd}G^{(TT;M^2)}_{cda'b'}(x,x')
& \equiv & L_{ab}^{({\rm inv}) cd}G^{(TT;M^2)}_{cda'b'}(x,x')\nonumber \\
&&  + \frac{1}{2}M^2
G_{aba'b'}^{(TT)}(x,x') - \frac{1}{2}M^2g_{ab}g^{cd}G_{cda'b'}^{(TT;M^2)}(x,x')\nonumber \\
  &  = &
\frac{1}{2} \left(-\Box + 2 + M^2\right)G^{(TT;M^2)}_{aba'b'}(x,x') \nonumber \\
&  = & \delta^{(TT)}_{aba'b'}(x,x'),  \label{GTTM}
\eeqa
where $L_{ab}^{({\rm inv}) cd}$ is defined by Eq.~(\ref{ffeq}). The transverse-traceless
delta-function is
\beq
\delta^{(TT)}_{aba'b'}(x,x') = \sum_{n=2}^\infty \sum_\nu K_{ab}^{(n\nu)}(x)
\overline{K_{a'b'}^{(n\nu)}(x')}.
\eeq
We clearly have
\beq
G^{(TT;M^2)}_{aba'b'}(x,x') = 2\sum_{n=2}^\infty \sum_\nu \frac{K_{ab}^{(n\nu)}(x)
\overline{K_{a'b'}^{(n\nu)}(x')}}{n(n+3)+M^2},
\eeq
and
\beq
G^{(TT)}_{aba'b'}(x,x') = \lim_{M\to 0} G^{(TT;M^2)}_{aba'b'}(x,x'),
\eeq

For spacelike-separated points $x$ and $x'$ the Lorentzian tensor two-point function $\Delta^{(TT;M^2)}_{aba'b'}(x,x')$ equals the Green's function on $S^4$ satisfying
the same equation as $G^{(TT;M^2)}_{aba'b'}(x,x')$, i.e.\ the first line of Eq.~(\ref{GTTM}),
but with the transverse-traceless delta-function
$\delta^{(TT)}_{aba'b'}(x,x')$ replaced by the full delta-function given by~\cite{AllenTuryn}
\beq
\delta_{aba'b'}(x,x') = \delta^{(TT)}_{aba'b'}(x,x')
+ \delta^{(TV)}_{aba'b'}(x,x') + \delta^{(TS)}_{aba'b'}(x,x'),
\eeq
where
\beq
\delta^{(TV)}_{aba'b'}(x,x') =  \sum_{n=2}^\infty\sum_{\nu} \frac{2\nabla_{(a} V^{(n\sigma)}_{b)}(x)
\overline{\nabla_{(a'}V^{(n\sigma)}_{b')}(x')}}{(n+1)(n+2)-6},
\eeq
and, with the definition $\lambda_n=n(n+3)$,
\beqa
\delta^{(TS)}_{aba'b'}(x,x') & = &\sum_{n=2}^\infty \sum_\nu
\frac{4}{3\lambda_n(\lambda_n-4)}
\left( \nabla_a\nabla_b + \frac{\lambda_n}{4} g_{ab}\right)\psi^{(n\nu)}(x)
\left(\nabla_{a'}\nabla_{b'} + \frac{\lambda_n}{4}g_{a'b'}\right)\overline{\psi^{(n\nu)}(x')}\nonumber \\
&& + \frac{1}{4}g_{ab}g_{a'b'}\sum_{n=0}^\infty \sum_\nu \psi^{(n\nu)}(x)\overline{\psi^{(n\nu)}(x')}.
\eeqa
One can find $\Delta_{aba'b'}^{(TT;M^2)}$ in the form
\beq
\Delta_{aba'b'}^{(TT;M^2)}(x,x')
= G_{aba'b'}^{(TT;M^2)}(x,x') + G_{aba'b'}^{(TV;M^2)}(x,x') + G_{aba'b'}^{(TS;M^2)}(x,x'), \label{relation}
\eeq
where
\beqa
L^{(M^2)cd}_{ab}G_{cda'b'}^{(TV)}(x,x') & = & \delta_{aba'b'}^{(TV)}(x,x'), \label{LM2TV}\\
L^{(M^2)cd}_{ab}G_{cda'b'}^{(TS)}(x,x') & = & \delta_{aba'b'}^{(TS)}(x,x'). \label{LM2TS}
\eeqa

By noting that
\beq
{L_{ab}}^{(M^2)cd}(\nabla_c V_d + \nabla_dV_c) = \frac{M^2}{2}(\nabla_aV_b + \nabla_bV_a)
\eeq
one can readily solve Eq.~(\ref{LM2TV}) as
\beqa
G_{aba'b'}^{(TV;M^2)}(x,x') & = & \frac{2}{M^2}\delta_{aba'b'}^{(TV)}(x,x') \nonumber \\
& = & \frac{2}{M^2}\lim_{\mu^2\to -6}\left[\nabla_{(a}\nabla_{|a'|}G^{(V;\mu^2)}_{b)b'}(x,x')
+ \nabla_{(a}\nabla_{|b'|}
G^{(V;\mu^2)}_{b)a'}(x,x')\right]\nonumber \\
& = & \frac{2}{M^2}\lim_{\mu^2\to -6}\left[\nabla_{(a}\nabla_{|a'|}\Delta^{(V;\mu^2)}_{b)b'}(x,x') + \nabla_{(a}\nabla_{|b'|}\Delta^{(V;\mu^2)}_{b)a'}(x,x')\right]\nonumber \\
&& + \frac{2}{3M^2}\nabla_a\nabla_b\nabla_{a'}\nabla_{b'}\Delta_0^{-}(x,x'). \label{GTV}
\eeqa

To find $G^{(TS;M^2)}_{aba'b'}(x,x')$ we first observe
\beqa
{L_{ab}}^{(M^2)cd}\nabla_c\nabla_d\psi^{(n\nu)} &=& \frac{M^2}{2}\nabla_a\nabla_b\psi^{(n\nu)} + \frac{M^2}{2}\lambda_n g_{ab}\psi^{(n\nu)},\\
{L_{ab}}^{(M^2) cd}g_{cd}\psi^{(n\nu)} & = & - \nabla_a\nabla_b\psi^{(n\nu)} - \left(\lambda_n - 3 + \tfrac{3}{2}M^2\right)g_{ab}\psi^{(n\nu)}.
\eeqa
The function $G^{(TS;M^2)}_{aba'b'}(x,x')$ can be found as the inverse of the operator
${L_{ab}}^{(M^2)cd}$
for the modes $g_{ab}\psi^{(n\nu)}$ and $(\nabla_a\nabla_b + \frac{\lambda_n}{4}g_{ab})\psi^{(n\nu)}$ as
\beqa
&& G^{(TS;M^2)}_{aba'b'}(x,x')\nonumber \\
 && =  -\frac{2}{3}\sum_{n=2}^\infty\sum_\nu\left( \frac{1}{M^2\lambda_n} + \frac{1}{(2-M^2)(\lambda_n-4)}\right)\nonumber \\
&& \,\,\,\,\,\times \left(\nabla_a\nabla_b + \frac{\lambda_n}{4}g_{ab}\right)\left(\nabla_{a'}\nabla_{b'} + \frac{\lambda_n}{4}g_{a'b'}\right)\psi^{(n\nu)}(x)
\overline{\psi^{(n\nu)}(x')} \nonumber \\
&& + \frac{1}{3M^2(2-M^2)}\nonumber \\
&& \times \sum_{n=2}^\infty\sum_\nu \left[g_{ab}\psi^{(n\nu)}(x)\left(\nabla_{a'}\nabla_{b'} + \frac{\lambda_n}{4}g_{a'b'}\right)\overline{\psi^{(n\nu)}(x')}
+ g_{a'b'}\overline{\psi^{(n\nu)}(x')}\left(\nabla_{a}\nabla_{b} + \frac{\lambda_n}{4}g_{ab}\right)\psi^{(n\nu)}(x)\right]\nonumber \\
&& + \sum_{n=0}^\infty\sum_\nu \frac{-\lambda_n + 2M^2}{12M^2(2-M^2)}g_{ab}g_{a'b'}\psi^{(n\nu)}(x)\overline{\psi^{(n\nu)}(x')}.
\eeqa
Some terms on the right-hand side have support only for $x=x'$ on $S^4$.  For example,
\beq
\sum_{n=0}^\infty\sum_\nu \frac{-\lambda_n + 2M^2}{M^2(2-M^2)}\psi^{(n\nu)}(x)\psi^{(n\nu)}(x')
= \frac{\Box + 2M^2}{M^2(2-M^2)}\delta(x,x').
\eeq
Thus, for $x\neq x'$ on $S^4$, or for spacelike-separated points $x$ and $x'$ in de~Sitter spacetime,
we have
\beqa
G^{(TS;M^2)}_{aba'b'}(x,x')
& = & - \frac{2}{3M^2}\nabla_a\nabla_b\nabla_{a'}\nabla_{b'}\Delta_0^{-}(x,x')\nonumber \\
&& - \frac{2}{3(2-M^2)}\left(\nabla_a\nabla_b - \frac{1}{4}g_{ab}\Box\right)\left(\nabla_{a'}\nabla_{b'} - \frac{1}{4}g_{a'b'}\Box'\right)
\Delta_{-4}^{--}(x,x'),\nonumber \\
\eeqa
where we have used the fact that $\Box \Delta^{-}_0(x,x')$ is a constant~\cite{HKcovariant}.
By substituting this equation and Eq.~(\ref{GTV}) into Eq.~(\ref{relation}) we find
\beqa
&& \Delta^{(TT;M^2)}_{aba'b'}(x,x') - G^{(TT;M^2)}_{aba'b'}(x,x')\nonumber \\
&& =\frac{2}{M^2}\lim_{\mu^2\to -6}\left[\nabla_{(a}\nabla_{|a'|}\Delta^{(V;\mu^2)}_{b)b'}(x,x') + \nabla_{(a}\nabla_{|b'|}\Delta^{(V;\mu^2)}_{b)a'}(x,x')\right]\nonumber \\
&& - \frac{2}{3(2-M^2)}\left(\nabla_a\nabla_b - \frac{1}{4}g_{ab}\Box\right)\left(\nabla_{a'}\nabla_{b'} - \frac{1}{4}g_{a'b'}\Box'\right)
\Delta_{-4}^{--}(x,x').
\eeqa
where $G^{(V;\mu^2)}_{aa'}(x,x')$ is defined by Eq.~(\ref{VecG}).
Then, noting that
\beqa
&& \lim_{M\to 0}\left\{\Delta^{(TT;M^2)}_{aba'b'}(x,x') - \frac{2}{M^2}\lim_{\mu^2\to -6}\left[\nabla_{(a}\nabla_{|a'|}\Delta^{(V;\mu^2)}_{b)b'}(x,x') + \nabla_{(a}\nabla_{|b'|}\Delta^{(V;\mu^2)}_{b)a'}(x,x')\right]\right\} \nonumber \\
&& = \sum_{\ell=2}^\infty\sum_\sigma H_{ab}^{(0;2\ell\sigma)}(x)\overline{H_{a'b'}^{(0;2\ell\sigma)}(x')} \nonumber \\
&& \,\,+ \lim_{M\to 0} \frac{1}{M^2}\sum_{m=0}^1 \sum_{\ell=2}^\infty\sum_\sigma (-1)^{m+1}
\left[ H_{ab}^{(M^2;m\ell\sigma)}(x)\overline{H_{a'b'}^{(M^2;m\ell\sigma)}(x')}
- H_{ab}^{(0;m\ell\sigma)}(x)\overline{H_{a'b'}^{(0;m\ell\sigma)}(x')}\right] \nonumber \\
&& = \Delta^{(TT)}_{aba'b'}(x,x'),
\eeqa
and using Eq.~(\ref{DeltaTT}), we indeed find Eq.~(\ref{TTconsistent}).

\end{document}